\documentclass[letterpaper,twocolumn,10pt]{article}

\usepackage{zhanggroup}
\usepackage{xspace}
\usepackage{amsmath,amsfonts}
\usepackage{graphicx}
\usepackage{textcomp}
\usepackage{epsfig}
\usepackage{url}
\usepackage{cite}
\usepackage{fancybox}
\usepackage{xcolor}
\usepackage{tikz}
\usepackage{amsmath}
\usepackage{eucal}
\usepackage{epsfig,endnotes}
\usepackage{pifont}
\usepackage[normalem]{ulem}
\usepackage{gensymb}
\usepackage{multirow}
\usepackage{subcaption}
\usepackage{caption}
\usepackage{mathtools}
\usepackage{bbm}
\usepackage{booktabs}
\usepackage{makecell}
\usepackage{amsmath}
\usepackage{enumitem}
\usepackage[absolute]{textpos}
\usepackage{multicol}
\usepackage{float}
\newcommand{\mypara}[1]{\noindent\textbf{#1.}}
\begin{document}
% ----------------------------------------------------

\begin{textblock}{12}(2,1)
\centering
To Appear in the 44th IEEE Symposium on Security and Privacy, May 22-25, 2023.
\end{textblock}

\title{\Large \bf On the Evolution of (Hateful) Memes by Means of Multimodal\\ Contrastive Learning\vspace*{-0.3cm}}

\author{
Yiting Qu\textsuperscript{\includegraphics[scale=0.1]{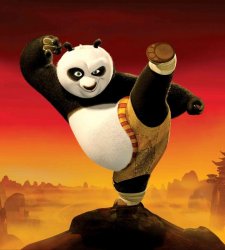}}\ \ \ \ \ 
Xinlei He\textsuperscript{\includegraphics[scale=0.1]{panda.jpeg}}\ \ \ \ \ 
Shannon Pierson\textsuperscript{\includegraphics[scale=0.1]{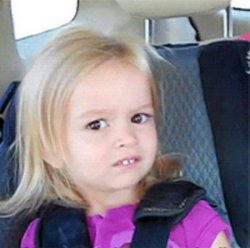}}\ \ \ \ 
Michael Backes\textsuperscript{\includegraphics[scale=0.1]{panda.jpeg}}\ \ \ \ \ 
\\
Yang Zhang\textsuperscript{\includegraphics[scale=0.1]{panda.jpeg}}\ \ \ \ \ 
Savvas Zannettou\textsuperscript{\includegraphics[scale=0.1]{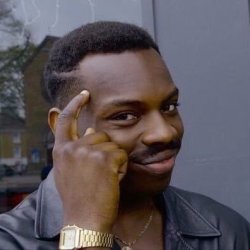}}
\\
\textsuperscript{\includegraphics[scale=0.1]{panda.jpeg}}\textit{CISPA Helmholtz Center for Information Security} \\
\textsuperscript{\includegraphics[scale=0.1]{meme3.jpeg}}\textit{London School of Economics and Political Science}\ \ \ \ \ \ 
\textsuperscript{\includegraphics[scale=0.1]{rollsafe.jpeg}}\textit{Delft University of Technology}
}

\date{
\textcolor{red}{
\small \textbf{Disclaimer:} This manuscript contains uncensored hateful content, such as antisemitic symbols that are highly offensive and might disturb the readers. We emphasize that we do not aim to propagate these hateful symbols. We include them in the paper to portray the nature and extent of hateful content online, raise awareness about these symbols, and provide the necessary contextual understanding of hateful connotations.}
}

\maketitle

% ----------------------------------------------------
\begin{abstract}
% ----------------------------------------------------

The dissemination of hateful memes online has adverse effects on social media platforms and the real world.
Detecting hateful memes is challenging, one of the reasons being the evolutionary nature of memes; new hateful memes can emerge by fusing hateful connotations with other cultural ideas or symbols.
In this paper, we propose a framework that leverages multimodal contrastive learning models, in particular OpenAI's CLIP, to identify targets of hateful content and systematically investigate the evolution of hateful memes.
We find that semantic regularities exist in CLIP-generated embeddings that describe semantic relationships within the same modality (images) or across modalities (images and text).
Leveraging this property, we study how hateful memes are created by combining visual elements from multiple images or fusing textual information with a hateful image.
We demonstrate the capabilities of our framework for analyzing the evolution of hateful memes by focusing on antisemitic memes, particularly the Happy Merchant meme.
Using our framework on a dataset extracted from 4chan, we find 3.3K variants of the Happy Merchant meme, with some linked to specific countries, persons, or organizations.
We envision that our framework can be used to aid human moderators by flagging new variants of hateful memes so that moderators can manually verify them and mitigate the problem of hateful content online.\footnote{Our code is available at \url{https://github.com/YitingQu/meme-evolution}.}

% ----------------------------------------------------
\end{abstract}
% ----------------------------------------------------

% ----------------------------------------------------
\section{Introduction}
\label{section: introduction}
% ----------------------------------------------------

Memes~\cite{D76,memes} are a popular way to communicate ideas across the Web, usually in text, images, or short videos.
In their simplest form, memes comprise a combination of visuals and text to disseminate an idea in a concise, engaging, and easily portable manner.
Generally, people share memes on the Web with benign intentions, e.g., being humorous or ironic.
However, memes can also be generated and spread for malicious purposes like coordinated hate campaigns~\cite{MSBCKLSS19}.
Fringe Web communities like 4chan~\cite{4chan} generate and disseminate many memes that have hateful connotations (e.g., antisemitic memes~\cite{ZFBB20}) or are politically charged~\cite{ZCBCSSS18}.
These memes can affect peoples' online experience and potentially lead to online radicalization~\cite{ROWAJ20,BS20} or even real-world hate crimes~\cite{wapo_violence}.
Given the likelihood of hateful memes causing real-world harm, there is a pressing need to detect and moderate instances of such memes.

\begin{figure}
\centering
\includegraphics[width=\columnwidth]{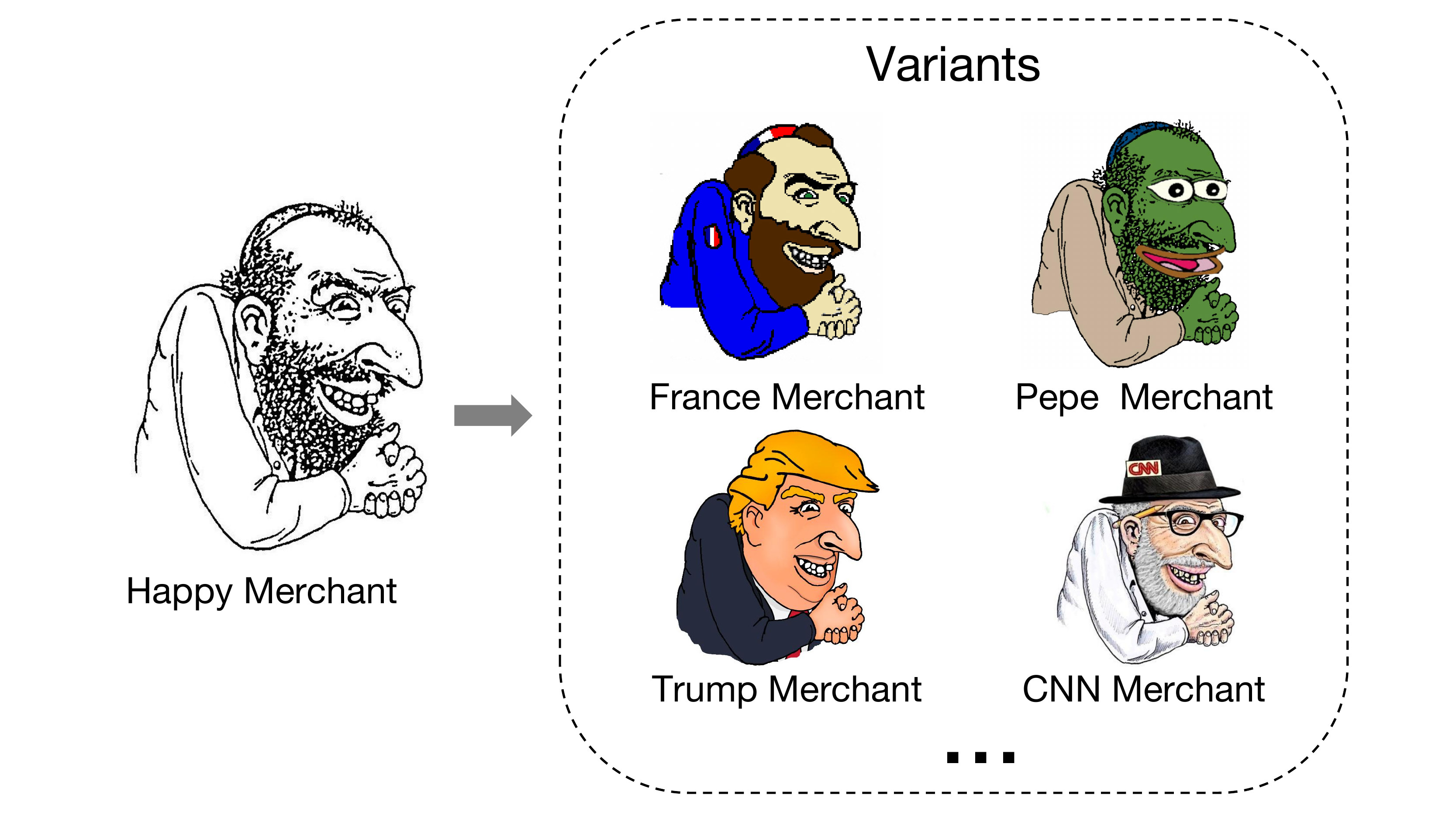}
\caption{Examples of Happy Merchant (a notorious antisemitic symbol) meme variants.}
\label{figure:variant_examples}
\end{figure}

Detecting and moderating hateful memes is a challenging task for several reasons.
First, memes encapsulate visual and textual information; hence it is challenging to capture the semantics of memes and identify whether memes share hateful connotations.
Second, memes have several features analogous to biological evolution~\cite{D76}, like variation, mutation, and inheritance.
Hateful memes constantly evolve as new memes can emerge by fusing other memes or cultural ideas.
For instance, considering the antisemitic Happy Merchant meme~\cite{happymerchant}, we can observe several variants in \autoref{figure:variant_examples} created because of other cultural ideas or symbols.
Memes' evolutionary nature makes detecting hateful memes even more challenging, as newly emerging memes will likely avoid detection from existing detection mechanisms.
For instance, Facebook relies on hashing techniques to identify near identical harmful content based on a database of already existing harmful images/videos~\cite{fb_hashing_moderation}.
However, this approach is incapable of dealing with the evolutionary nature of hateful memes (images can share hateful connotations and have substantially different hashes, hence remaining undetected).
Taken altogether, these challenges highlight the need for designing automated tools/techniques that identify the variants and the evolution of hateful memes, as well as identifying the main themes or cultural symbols causing the creation of many hateful memes.

In this paper, we contribute to detecting and understanding hateful memes' evolution using state-of-the-art Artificial Intelligence (AI) models.
We use AI models that use the contrastive learning paradigm, specifically OpenAI's Contrastive Language–Image Pre-training (CLIP) model~\cite{RKHRGASAMCKS21}, to design and implement a framework that allows us to identify the main targets of hateful memes and systematically analyze the evolution of memes.
The CLIP model embeds text and images into the same vector space, allowing us to assess semantic similarities and extract relationships between textual and image-based features.
In particular, CLIP can serve as an image or text search engine given a specified query, which enables us to retrieve the most relevant image or text based on the input and the dataset.
Also, we find and use another property of CLIP, semantic regularities, which describe that image and text embeddings capture semantic relationships that can be transferred within modalities (image to image) or across modalities (text to image) via algebraic operations on embeddings like summation and subtraction.

Using these two CLIP advantages, we implement a framework for identifying hateful content's main themes and targets.
We use the CLIP model to embed contents (memes and language contexts) into the high-dimensional vector space; then, we perform clustering, and automatic annotation of clusters, including whether clusters are used in hateful contexts.
Also, we systematically analyze the evolution of hateful memes by incorporating CLIP's semantic regularities in our framework.
To the best of our knowledge, our work is the first one that systematically discovers and uses semantic regularities that exist on CLIP embeddings to study the problem of hateful content on the Web.
We analyze the evolution of hateful memes using two semantic regularities; semantics are transferred within images (visual semantic regularities) and across images and texts (visual-linguistic semantic regularities).
The former aims to identify hateful meme variants and how other images influence them.
The latter aims to identify hateful meme variants based on a set of pre-defined named entities (e.g., countries, persons, etc.).
We validate the efficacy of our proposed framework about meme evolution by focusing on antisemitic hateful memes, particularly the Happy Merchant meme.

In general, our contributions can be summarized as the following:
\begin{itemize}
\item We propose a framework that can automatically capture and fuse the rich semantics of memes (both visual features and language context).
The framework can identify the main groups of potentially hateful memes in an unsupervised manner using clustering and hate measurement techniques.
Our approach extends previous efforts in identifying targets of hateful content mainly because it fuses text and image modalities on content shared on social media.
It can also assist content moderators in understanding the targets of hateful memes holistically and mitigate the effects of spreading this harmful content.
\item We provide an automated and scalable method that leverages CLIP's semantic regularities to identify meme variants and potential influencers to understand hateful memes' creation, variation, and evolution.
We argue that this framework can provide novel insights related to the creation and evolution of memes on social media platforms (e.g., the mutation rate and breath of hateful memes, the lifespan of hateful meme variants, etc.).
This framework can be paramount for researchers or social media operators working on tackling emerging socio-technical issues like hateful content, as our framework can effectively identify images that fuse semantics from images/text to spread harmful content.
We believe that future work should explore the possibility of using our framework for detecting other potentially harmful information like misinformation images.
\item We make our framework publicly available, allowing researchers to study the evolution of other hateful memes.
Additionally, we will make all the discovered Happy Merchant variants available upon request (to avoid malicious uses of the dataset, such as the automatic generation of hateful memes using Generative Adversarial Networks), which we believe will provide a valuable dataset to researchers working on antisemitism and social media operators aiming to limit the spread of antisemitic memes.
\end{itemize}

\mypara{Ethical considerations} 
Our work analyzes publicly available anonymous datasets from 4chan's /pol/ board.
We emphasize that our work performs passive measurements by analyzing the content shared by anonymous users on 4chan.
We follow standard ethical guidelines when analyzing the data and presenting the results, including reporting results on aggregate, protecting the anonymity/privacy of the users, and not attempting to track users across websites~\cite{RL14}.

Also, we believe that it is important from an ethical point of view to clarify why we elect to include hateful meme variants in this manuscript.
We emphasize that our goal is not to propagate and promote these hateful meme variants.
While we acknowledge that these symbols are likely disturbing for many readers, we believe that there are many advantages in showing them that outweigh the potential harm.
First, by demonstrating the various instances of hateful meme variants, we genuinely portray the nature and extent of hateful content online, which allows the readers to understand the severity and peculiarities of the problem fully.
Second, including these symbols assists in raising awareness of hate symbols, which is essential to understanding hateful content and its impact.
Third, we believe that including hateful meme variants provides the necessary contextual understanding that assists the readers in comprehending the applicability of our methods in tackling hateful content and the various connotations of hateful content across contexts, target groups, or geographical location. 

Our work focuses on the evolution of hateful memes and makes available the identified hateful meme variants.
We acknowledge the concerns about their potential misuse, e.g., a malicious user might leverage the identified hateful memes to train a generative model and scale the generation of hateful content.
To minimize the risk, we only provide the complete set of identified hateful memes upon request for research purposes.
Furthermore, all manual annotations are performed by the authors of this paper, hence there is no exposure of third parties to the hateful memes.

% ----------------------------------------------------
\section{Background}
\label{section: background}
% ----------------------------------------------------

This section provides background information on 4chan and our dataset, as well as an overview of the CLIP model.

\mypara{4chan dataset}
4chan~\cite{4chan} is an anonymous image board known for creating and disseminating a substantial number of Internet memes.
Due to the anonymity of users and lack of moderation, 4chan is the subject of media attention relating to far-right and neo-Nazi ideology~\cite{BMHAPV11, ZFBB20, MZBC20}.
4chan is divided into sub-communities called boards, each with a specific topic of interest.
In this work, we focus on 4chan's ``Politically Incorrect'' board (/pol/), which focuses on discussing world events and politics.
Viral conspiracy theories and toxic memes often originate in fringe online communities like 4chan’s /pol/ and migrate to and proliferate on mainstream platforms~\cite{C12}.
Studies~\cite{ZCBCSSS18, ZFBB20} showed that /pol/ is particularly influential in propagating racist/political memes and conspiracy theories into other online communities.
To study the spread and evolution of memes on 4chan's /pol/, we use a dataset collected by Zannettou et al.~\cite{ZCBCSSS18} that includes all images (4.3M) shared on /pol/ posts from June 30, 2016, to July 31, 2017.
We complement this dataset with information about the text of the posts using the dataset released by Papasavva et al.~\cite{PZCSB20}.
In particular, we filter all posts that include any of the 4.3M images, hence obtaining a set of 12.5M image-text pairs.

We choose this dataset for two reasons: 1)~Since we want to demonstrate the applicability of our framework in detecting hateful memes, and in particular, identifying many variants of the same hateful meme, we select a fringe Web community that is known for the dissemination and creation of a large number of hateful memes~\cite{ZCBCSSS18, ZFBB20};
2)~The same dataset is used in previous work to study online hate speech~\cite{GZ22, ZFBB20}.
Also, given that the CLIP model has great generalizability and can be applied for various downstream tasks, we expect that our framework can also be applied to other datasets beyond our 4chan's /pol/ dataset.

\mypara{OpenAI's CLIP}
OpenAI's Contrastive Language-Image Pre-training (CLIP)~\cite{RKHRGASAMCKS21} is a novel approach using natural language supervision for image representation learning.
Conventional contrastive learning techniques like SimCLR~\cite{CKNH20}, BYOL~\cite{GSATRBDPGAPKMV20}, and MoCo~\cite{CFGH20} utilize data augmentation to train image encoders in a self-supervised manner.
Differently, CLIP jointly trains an image encoder and a text encoder to predict the correct image-text pairs instead of image-image pairs.
The model learns visual and linguistic embeddings simultaneously by minimizing the cosine similarity of the image and text embeddings from the same pair.
The training data also does not require manual labeling because texts in the dataset provide supervision in the text-image contrastive training.
The availability of a vast amount of data online that contains both images and texts, such as articles and posts, has made large-scale contrastive training practical.
OpenAI constructed a dataset of 400 million (image-text) pairs collected online from various publicly available sources.
The CLIP model learns the general representation of both image and text and connects both modalities, which enables zero-shot transfer learning.

\begin{figure}
\centering
\includegraphics[width=\columnwidth]{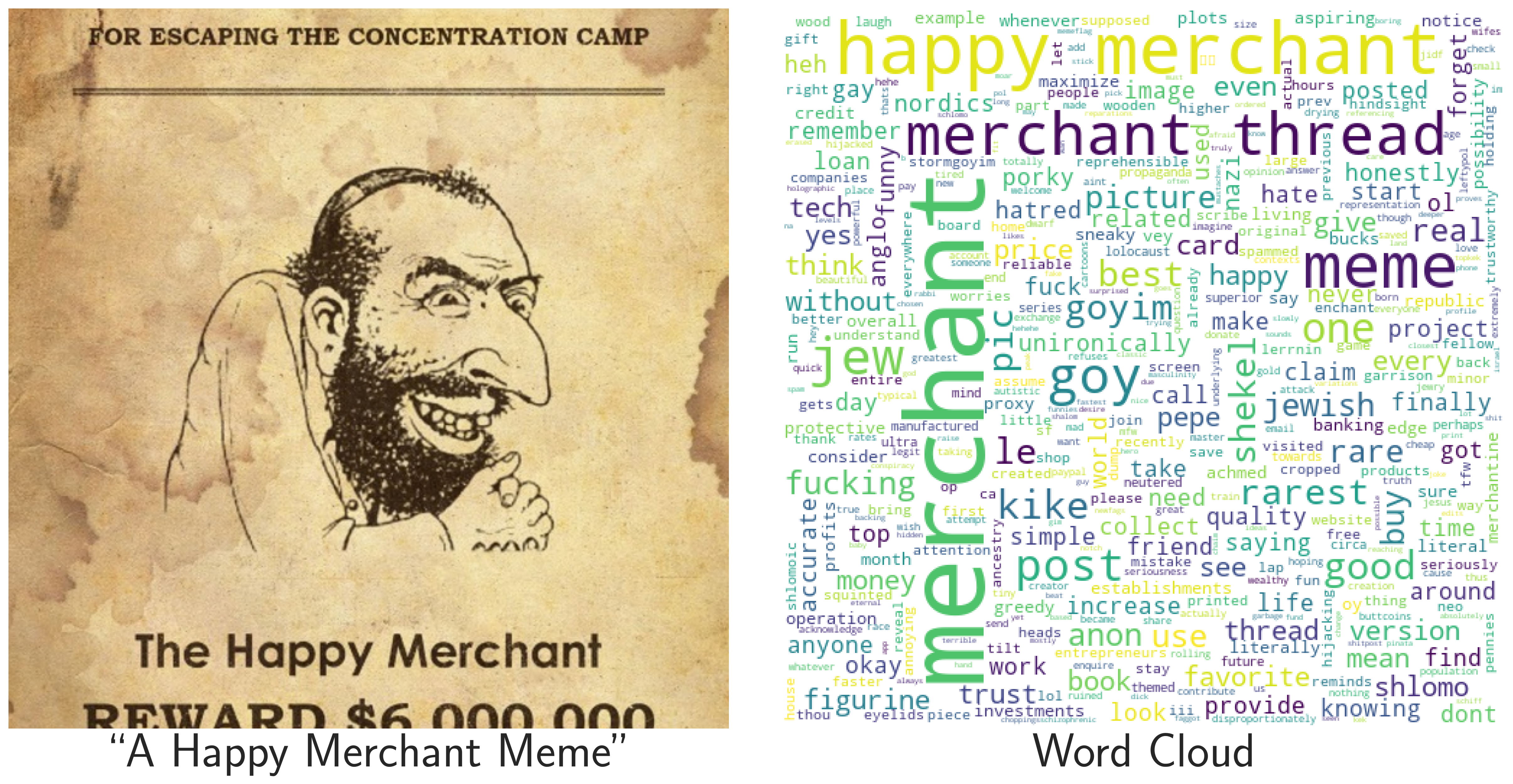}
\caption{Example of retrieved image/text using CLIP. We use ``A Happy Merchant Meme'' (the name of the antisemitic signal) as the initial query.}
\label{figure:retrieval}
\end{figure}

% ----------------------------------------------------
\section{Using CLIP for Online Hate}
\label{section: clip_targeting_online_hate}
% ----------------------------------------------------

% ----------------------------------------------------
\subsection{Fine-tuning CLIP}
\label{subsection: fine_tuning_clip}
% ----------------------------------------------------

Even though OpenAI's pre-trained CLIP model has great generalizability of image and text representations~\cite{RKHRGASAMCKS21}, fine-tuning CLIP on the 4chan dataset is helpful for the following reasons.
First, OpenAI did not disclose the exact methodology for creating the 400M text-image pairs used for training the model.
However, given the nature of the platform (i.e., dissemination of fringe or hateful ideologies), we expect that 4chan-related activity is not included in CLIP's training data.
Second, different from general datasets obtained from the Web, 4chan's data is filled with slurs and 4chan-specific slang language, likely not effectively captured by the pre-trained CLIP model.
Take the meme's name, for example, Feels Good Man~\cite{feelsgoodman} is a typical catchphrase in daily life; in 4chan, however, it represents a popular meme of a smiling frog.
The fine-tuned CLIP is expected to better connect such phrases and their related images.
Finally, we expect the fine-tuned CLIP to build a connection between popular image memes and 4chan-related topics.
For example, Pepe the Frog~\cite{pepethefrog} might have been learned by the pre-trained CLIP model as a popular meme.
However, in 4chan, Pepe the Frog is often closely used in discussions relating to  Jews, Muslims, and famous politicians; we expect that such peculiarities will be hard to capture by the pre-trained CLIP model.
In this work, we fine-tune the entire CLIP model (ViT-B/32), including the image encoder, text encoder, and the final projection layers, using 10.4 million image-text pairs as the training set and 2.1 million pairs as the testing set.
We provide more fine-tuning details and evaluation in \autoref{appendix:finetuning} in Appendix, where we also show that the fine-tuned CLIP performs better than the pre-trained model in recognizing the same image-text pair in the 4chan dataset.

% ----------------------------------------------------
\subsection{CLIP's Versatile Applications} 
\label{subsection: flexible applications}
% ----------------------------------------------------

Here, we describe how we use CLIP to analyze our 4chan dataset in the wild.
The multimodal embeddings extracted from CLIP are the foundation for various downstream tasks.
This paper uses the image and text embeddings from CLIP's final projection layers as multimodal representations.
We use the term \emph{embeddings} to refer to CLIP's image/text representations for the rest of the paper.

\mypara{Image \& text retrieval}
We can utilize CLIP as a search engine for retrieving relevant images/text (i.e., similar text/images based on the CLIP embeddings), given a query that is either text or an image.
To demonstrate this application, we create a dataset by randomly selecting 1M image/text pairs shared on 4chan's /pol/ (out of 12.5M posts).
\autoref{figure:retrieval} shows two examples.
We use the query ``A Happy Merchant Meme'' and find the image with the embedding with the largest cosine similarity to the embedding extracted using the text query.
The resulting image is the Happy Merchant meme, demonstrating that CLIP identifies the meme.
Next, we use the same image for text retrieval and collect the top-100 textual posts in terms of the cosine similarity between the image and text embeddings.
A word cloud is created based on the top-100 textual posts.

\begin{figure}[t]
\centering
\begin{subfigure}{\columnwidth}
\centering
\includegraphics[width=\columnwidth]{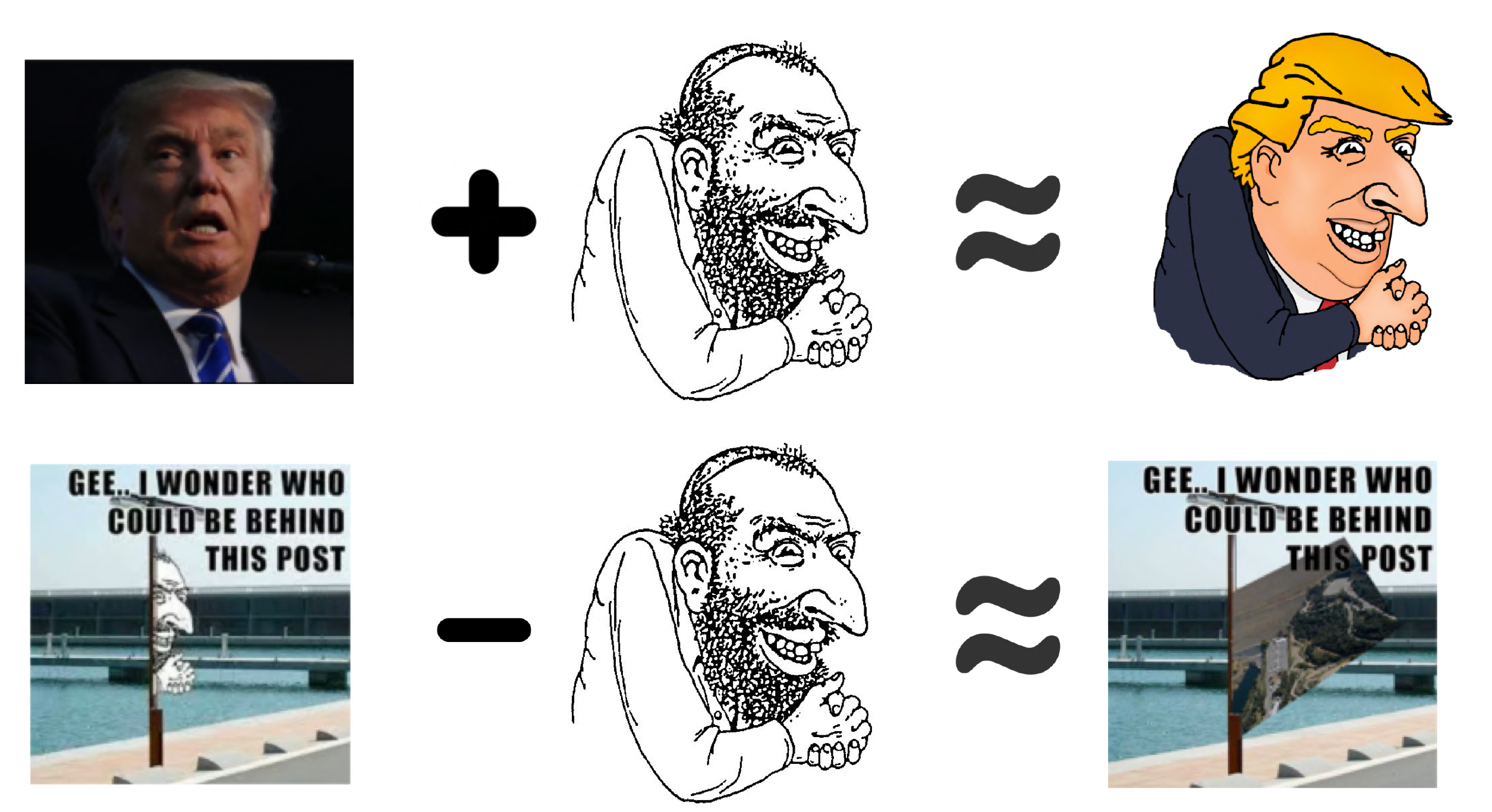}
\caption{Visual semantic regularities}
\label{figure:manipulation_0}
\end{subfigure}
\begin{subfigure}{\columnwidth}
\centering
\includegraphics[width=\columnwidth]{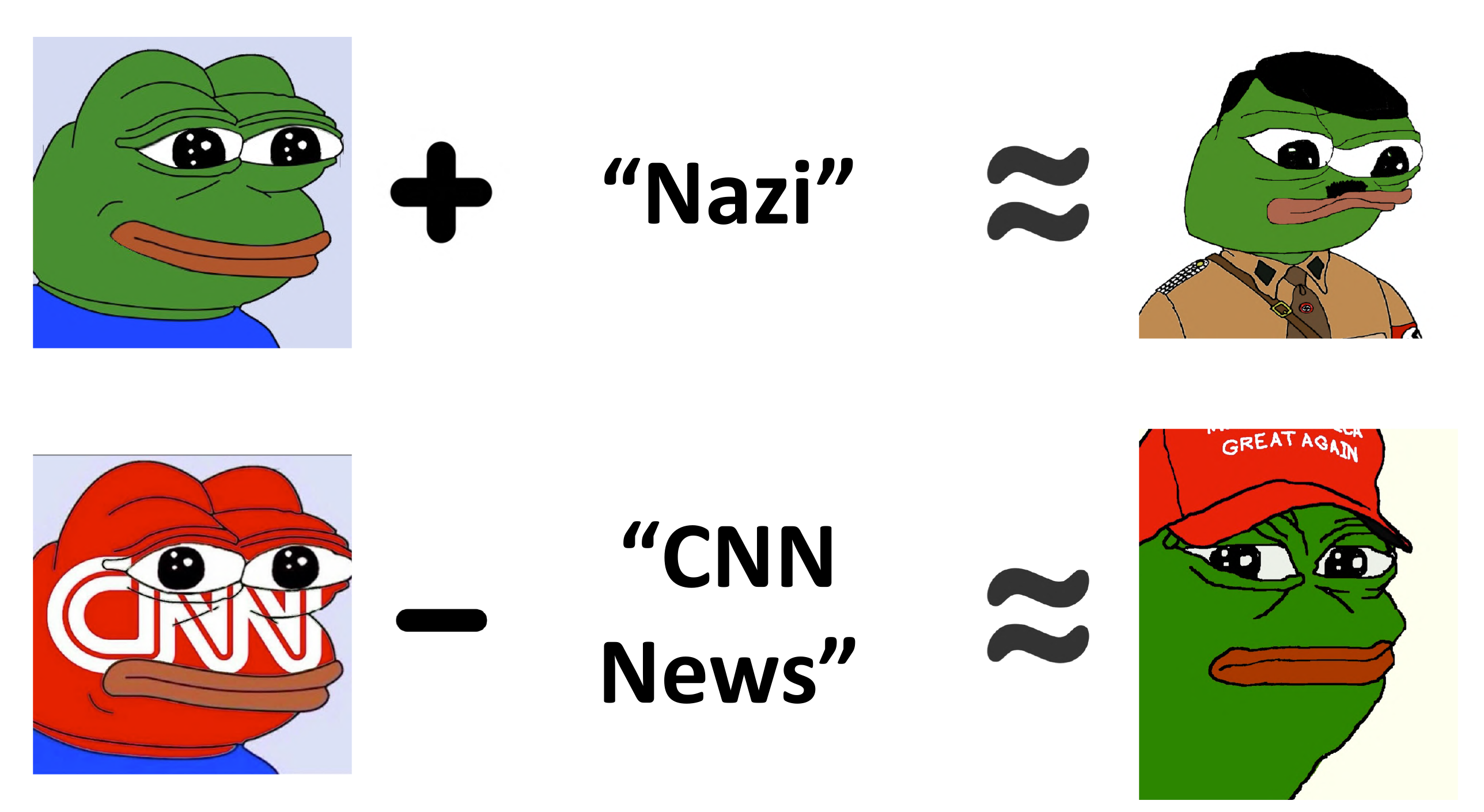}
\caption{Visual-linguistic semantic regularities}
\label{figure:manipulation_1}
\end{subfigure}
\caption{Examples of semantic regularities. We show instances of hateful memes, including Happy Merchant and Pepe the Frog, to illustrate how semantic regularities can aid in the study of the evolution of such memes.}
\label{figure:manipulations}
\end{figure}

\mypara{Demonstrating semantic regularities} 
CLIP embeddings encapsulate semantic relationships, similarly to word vectors from Word2vec~\cite{MCCD13} models.
With simple algebraic operations on word vectors, e.g., $\textit{vector\{King\}}-\textit{vector\{Man\}}+\textit{vector\{Woman\}}$, the resulting vector will be close to $\textit{vector\{Queen\}}$.
Such properties on Word2vec are generally referred to as \emph{linguistic semantic regularities}~\cite{CCP20,GDM16}.
We observe that semantic regularities also exist in CLIP embeddings (see \autoref{appendix:probing} in Appendix to find the underlying reason for the existence of semantic regularities), with the difference that they can be observed across multiple modalities (i.e., text and images).
We group the semantic regularities into visual semantic regularities and visual-linguistic semantic regularities based on the modalities when performing operations, as introduced in the following.
To our knowledge, we conduct the first work that uses such properties in CLIP embeddings for studying online hate.

\emph{Visual semantic regularities} describe the semantic relations presented in images.
As shown in \autoref{figure:manipulation_0}, we perform algebraic operations on image embeddings.
Given an image of Donald Trump and an image of the Happy Merchant meme, we sum their embeddings with the same weights (0.5, 0.5) and search for the most similar image in the embedding space (i.e., the image that is closest to the embedding obtained after the summation).
The summation leads to an image combining elements from both images, demonstrating semantic regularities.
Similarly, we can extract semantic regularities by performing other operations such as subtraction (see the second example in \autoref{figure:manipulation_0}).

\emph{Visual-linguistic semantic regularities} describe similar relationships across different modalities.
We perform operations across image and text embeddings (see \autoref{figure:manipulation_1}).
For instance, given a Pepe the Frog image, we perform the summation operation on this image embedding and the embedding extracted from the text ``Nazi.''
For summation across modalities, we use 0.2 and 0.8 as the weights for image and text embeddings, respectively.
We choose the weights based on manual examinations, where we select ten image-text pairs for the summation operation.
We increase the weight of text embeddings from 0.5 to 0.9 with a step of 0.1 and observe that text often exerts limited influence on the final image until the weight reaches 0.8.
To identify the resulting image (right-hand images in \autoref{figure:manipulation_1}), we search for the closest image embedding to the summation embedding, resulting in an image from our dataset that has both visual features (frog) and linguistic features (Nazi).
Notice that no image or text generators are employed; we retrieve images from our 4chan dataset to validate the observed property.

We formalize the visual semantic regularities as follows:
considering 3 memes $m_A, m_B, m_C$ with embeddings $e_A, e_B, e_C$, if $\alpha * e_A + (1-\alpha)*e_B \approx e_C$, then visually, we might also have $m_A + m_B \approx m_C$.
$m_C$ will generally preserve the semantics/visual features from meme $m_A$ and $m_B$.
The fraction of both semantics depends on the weight $\alpha$.
We generalize this formulation to the visual-linguistic semantic regularities when $e_A, e_B, e_C$ represent either image or text embeddings, e.g., $e_A$ is the text embedding, and $e_B, e_C$ are image embeddings.
The summation result represented by $e_C$ can still preserve the semantics from the text embedding $e_A$ and the visual embedding $e_B$.
We aim to systematically extract semantic regularities from CLIP embeddings to study the evolution of hateful memes on 4chan (see \autoref{section: evolution}).

\begin{table*}[!t]
\centering
\caption{Statistics and annotation accuracy of clusters based on different embeddings.}
\setlength{\tabcolsep}{0.5em}
\scalebox{1.0}{
\begin{tabular}{lcccccccc}
\toprule
\multirow{2}{0.4em}{Clustering Embedding} &  \%Noise &\%Clustered &\#Clusters & \#Clusters  &  KeyBERT-V & KeyBERT-N & TextRank & Agreement\\
& & Posts & ($>$30) & & & & \\
\midrule
Image              & 48.3\% & 51.7\% & 26,618 & 1,901 & 0.95 & 0.95 & 0.95 &  0.91 \\
Text               & 47.9\% & 52.1\% & 1,522  & 116   & - & - & - & - \\
Fused (Image+Text) & 62.2\% & 37.8\% & 14,553 & 1,229 & 0.97 & 0.97 & 0.96 & 0.95 \\
\bottomrule
\end{tabular}
}
\label{table:cluster_info}
\end{table*}

\begin{figure*}[!t]
\centering
\begin{subfigure}{0.65\columnwidth}
\includegraphics[width=\columnwidth]{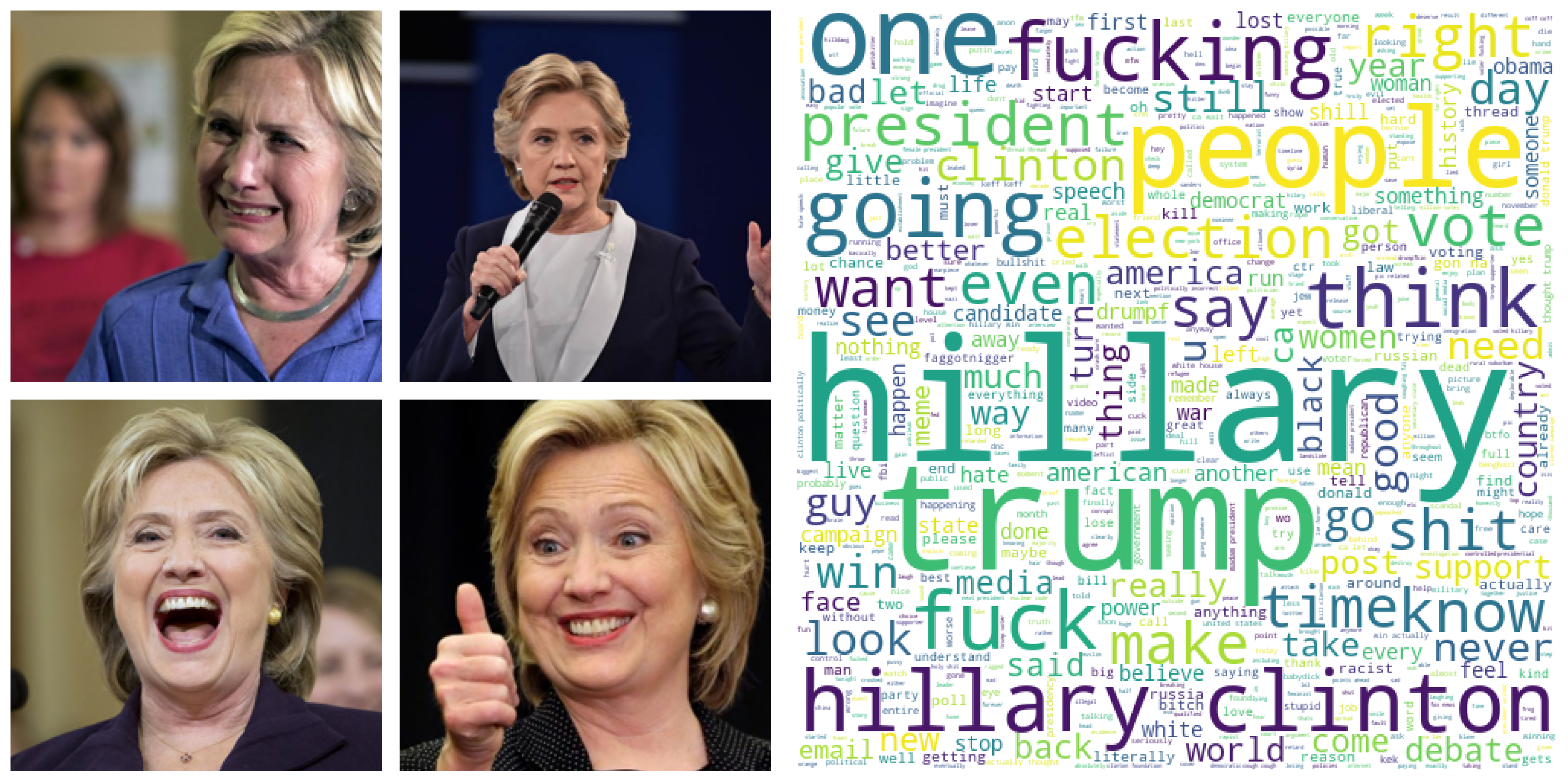}
\caption{\centering image embeddings; ``poll-shock-hillary''}
\label{figure:cluster_image}
\end{subfigure}
\begin{subfigure}{0.65\columnwidth}
\includegraphics[width=\columnwidth]{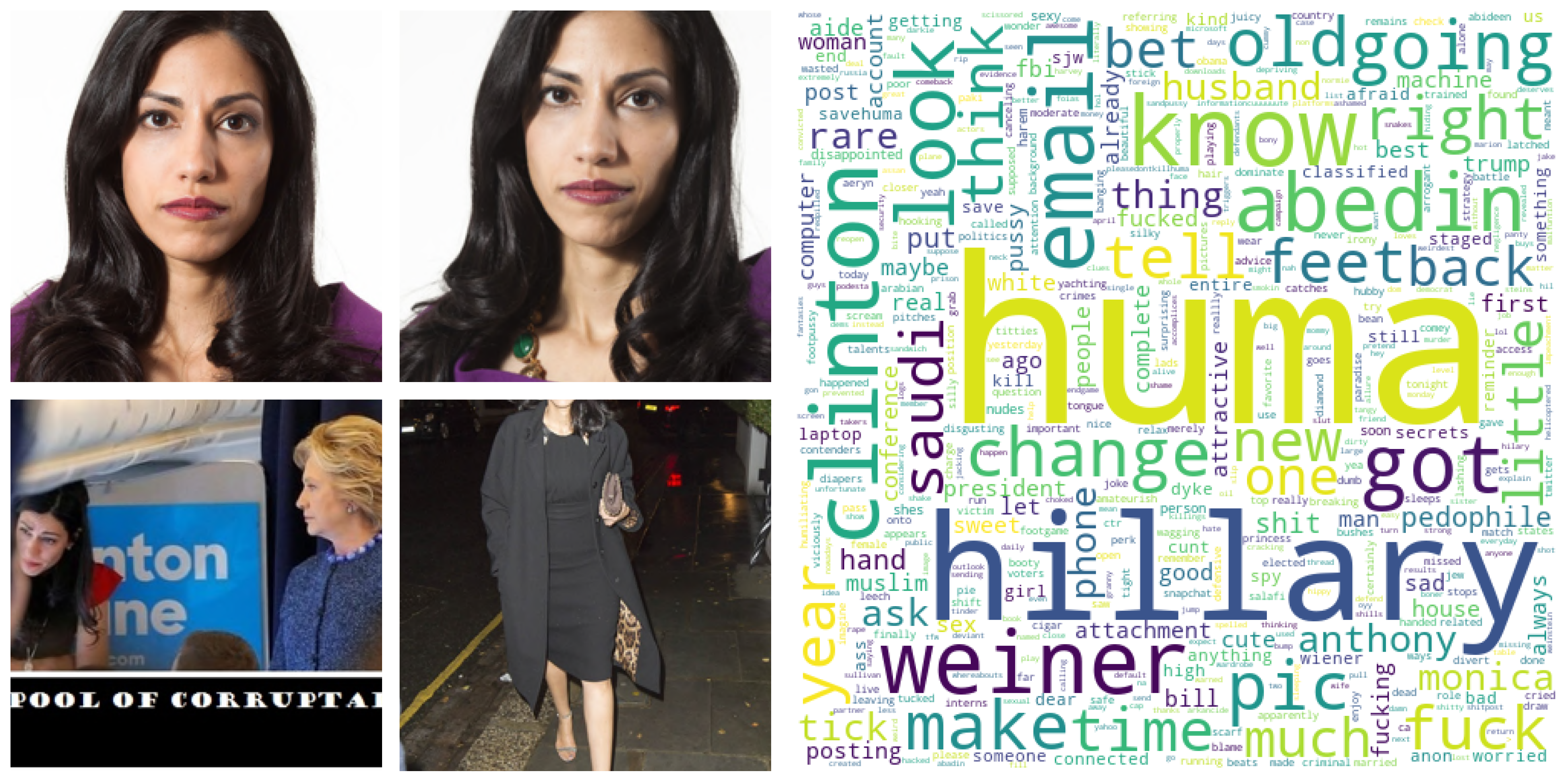}
\caption{\centering text embeddings; ``huma-remains-hillary''}
\label{figure:cluster_text}
\end{subfigure}
\begin{subfigure}{0.65\columnwidth}
\includegraphics[width=\columnwidth]{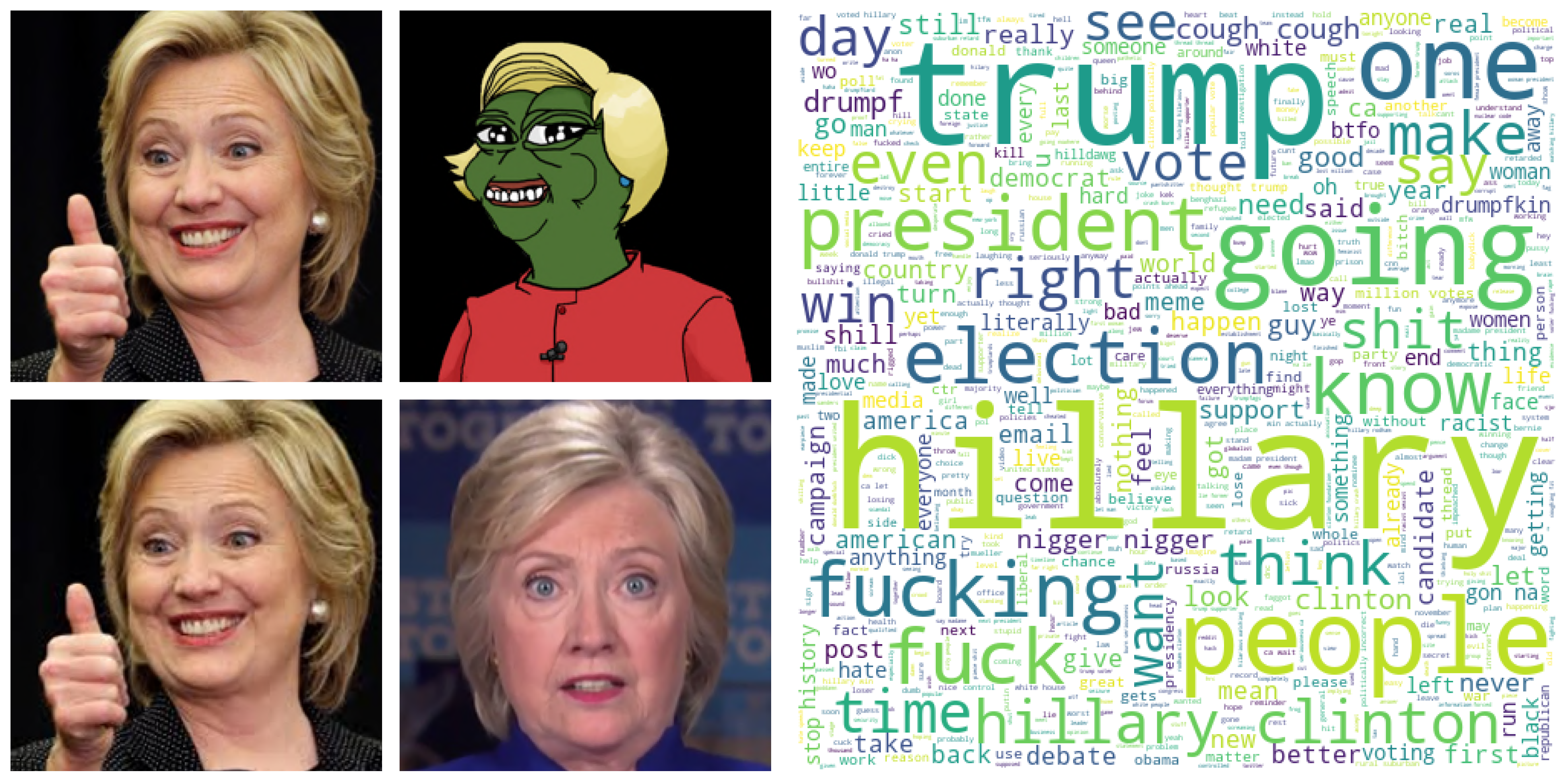}
\caption{\centering fused embeddings; ``president-hillary-running'' 
}
 \label{figure:cluster_sum}
\end{subfigure}
\caption{Examples from the cluster ``Hillary Clinton'' on different types of embeddings.
We randomly select 4 images from each cluster, and we visualize high-frequency words by building a word cloud with all textual comments in the cluster.}
\label{figure:clusters_res}
\end{figure*}

% ----------------------------------------------------
\section{Understanding Hateful Meme Clusters}
\label{section: clustering}
% ----------------------------------------------------

In this section, we propose a framework to understand, interpret, and assess the hate of memes in the textual context.
The conventional way to do this is either utilizing clustering techniques on images to form image clusters or topic modeling the text into several topics.
However, dealing with the single modality (building image clusters or modeling text topics) in an isolated manner hardly precisely describe the semantics of memes in different contexts.
In 4chan, the posted meme and the comment do not necessarily present the same semantics.
For example, a user comment on the picture of a politician with ``good job!'' without mentioning his name.
Processing such information from either side fails to bridge the gap in semantics from different modalities.

Motivated by CLIP's ability to process multimodal information, we creatively construct a new meme embedding space containing multimodal semantics by fusing visual and contextual embeddings.
Using the embeddings, we build meme clusters (\autoref{subsection: dbscan_clustering}), annotate meme clusters with key phrases (\autoref{subsection: annotation}), and finally perform a hate assessment (\autoref{subsection: hate_analysis}) to extract the main targets of hateful content.
Here, we randomly select 1M image-text pairs out of 12.5M in the dataset, including 1M comments and 0.5M unique images.
We then use the fine-tuned CLIP to obtain image and text embeddings (512-dimensional vectors).
Note that we adopt the random sampling strategy instead of using the entire dataset because the subset has a similar distribution with the entire dataset, and performing clustering on the subset significantly reduces computation time.

% ----------------------------------------------------
\subsection{Clustering}
\label{subsection: dbscan_clustering}
% ----------------------------------------------------

We construct the new meme embedding by summing the meme and contextual embedding, as embedding summation is verified to be an effective way to fuse semantics due to the visual-linguistic semantic regularities as introduced in \autoref{subsection: flexible applications}.
To compare the fused embedding clustering and the conventional single modality clustering, we perform clustering on three types of embeddings:
1)~Image embeddings: We focus on categorizing images by clustering the image embeddings.
2)~Text embeddings: Clustering on the text embeddings helps identify popular topics.
3)~Fused embeddings (image + text): For each image-text pair, we sum the image and text embedding to obtain the fused embedding that connects both semantics.

Inspired by previous works~\cite{ZCBCSSS18, ZFBB20}, we employ the Density-based spatial clustering of applications with noise (DBSCAN)~\cite{EKSX96} to build clusters.
DBSCAN separates clusters based on density and automatically infers the number of clusters.
In addition, it can detect irregular shapes clusters and is robust to outliers.
This advantage is apparent in 4chan data because many noise images and nonsense comments should be considered outliers.
DBSCAN relies on two parameters: $min\_samples$ and $eps$, which indicate the necessary density to form a cluster.
DBSCAN defines a core sample as a sample that has at least $min\_samples$ neighbors within a distance of $eps$.

Noise data are the outliers that do not belong to any cluster due to relatively larger distances.
We use Euclidean Distance as the distance metric and carefully tune the parameters based on different types of embeddings.
\autoref{table:cluster_info} reports the statistics of different clustering results.
Notice that there is no effective metric to evaluate the clustering performance on million-level data with substantial noise.
We determine the final DBSCAN parameters based on manual evaluations of the cluster quality, as well as the noise level and concentration.
All the noise levels shown in \autoref{table:cluster_info} are in the range of 47.9\%-62.2\%, consistent with the noise levels in~\cite{ZCBCSSS18}.
And the concentration of each cluster represents how likely all images or texts within the same cluster are concentrated on the same theme, semantic-wise.
We also manually check the top-50 clusters by randomly viewing members to avoid a high false positive rate (i.e., the ratio of samples that should not be part of the cluster).

\begin{figure*}[t]
\centering
\begin{subfigure}{\columnwidth}
\includegraphics[width=\columnwidth]{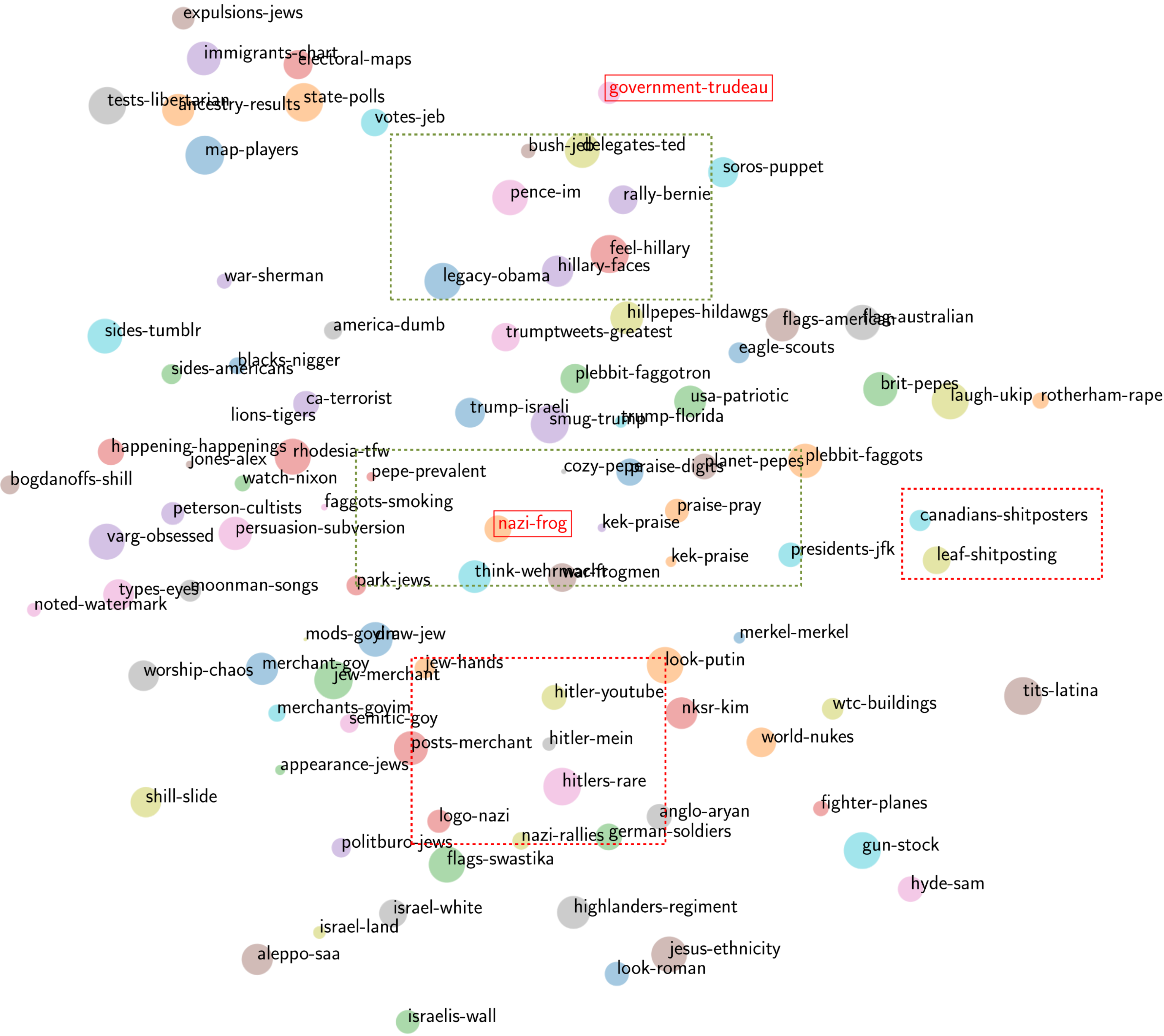}
\caption{Image embeddings}
\label{figure:annotation_image}
\end{subfigure}
\begin{subfigure}{\columnwidth}
\includegraphics[width=\columnwidth]{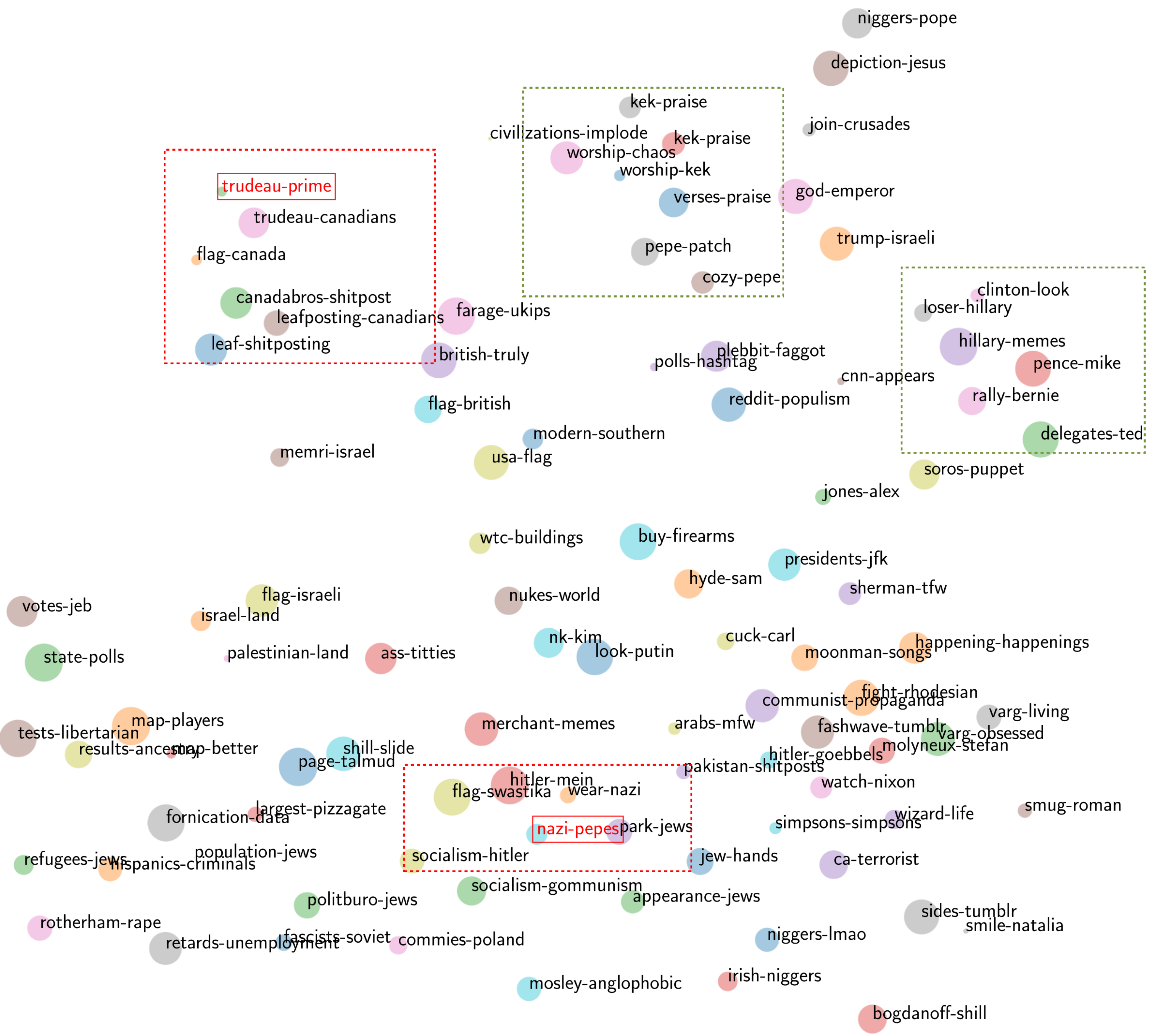}
\caption{Fused embeddings}
\label{figure:annotation_sum}
\end{subfigure}
\caption{t-SNE projections of top-100 centroid embeddings. Each node is a cluster, and the node is sized by the number of cluster samples. 
We mark the commonalities with green boxes and highlight the disparities with red boxes.}
\label{figure:annotation}
\end{figure*}

\mypara{Findings}
All three clustering results capture the dominated clusters, e.g., Comics, Beauties, Donald Trump, US Election, Nazi Ideology, Happy Merchants, etc.
Differently, each clustering presents specific patterns.
As illustrated in \autoref{figure:clusters_res}, image embedding clustering identifies clusters merely by the visual features.
Thus, the images within each cluster are highly similar.
Conversely, text embedding clustering relies on the comments and completely neglects the images.
The observed clusters present a high concentration text-wise; meanwhile, the images in the same cluster are sometimes irrelevant.
As a combined strategy, fused embedding clustering recognizes the images of common semantics into the same cluster, despite the apparent differences in visual or textual features.
Take the Hillary cluster as an example, image embedding clustering only contains the figure of Hillary, and text embedding clustering includes irrelevant images.
Still, fused embedding clustering can have images that are visually different but highly relevant in semantics.
This is an advantage compared to image embedding clustering in understanding 4chan's millions of memes and their semantics in the specific context.

% ----------------------------------------------------
\subsection{Automatic Cluster Annotation}
\label{subsection: annotation}
% ----------------------------------------------------

We aim to interpret the semantics of meme clusters explicitly with natural language.
Due to a large number of meme clusters (26,618 clusters in image embedding clustering), it is challenging to perform a manual inspection on all clusters.
To address this challenge, we employ CLIP as a search engine to retrieve similar sentences given an image embedding and then extract key phrases from the sentences to annotate the clusters with 2-3 words.

\mypara{Pipeline}
We first discard all the clusters that contain less than 30 samples.
For each remaining cluster, we compute its centroid embedding by averaging all the embeddings in the cluster.
Note that the centroid embedding here is averaged image embedding for image embedding clustering and multimodal embedding (image + text) for fused embedding clustering.
With the centroid embedding of each cluster, we then retrieve the top-300 most similar textual posts in the 1M image-text pairs by computing the cosine similarities, which serves as the document for key phrase extraction later.
After cleaning the collected document, e.g., removing stop words and non-alphas, we apply five types of key phrase extraction techniques on every cluster in the image and fused-based clustering results.
The reason to extract key phrases instead of keywords is to summarize the meme clusters better and maintain coherence, e.g., Pepe the Frog is a better label than Frog/Pepe.
The key phrases extraction methods include KeyBert (vectorizer)~\cite{keybert}, KeyBert (ngram)~\cite{keybert}, TextRank~\cite{pytextrank}, Yake~\cite{yake}, and Rake~\cite{rake}.
We only use KeyBert (vectorizer), KeyBert (ngram), and Textrank for evaluation due to the poor key phrase quality extracted by Yake and Rake.
We randomly select 50 clusters in each type of clustering to evaluate the annotation quality.
Every selected key phrases extraction method generates three candidates for final selection.
If all three candidates describe or interpret the contents of that cluster correctly, we then agree that the annotation is correct.
The evaluation process is manually conducted by two of the authors of this paper independently.
We extract key phrases for clusters in the image- and fused-based clustering.

As \autoref{table:cluster_info} shows, key phrase extractions present good annotating accuracy with a considerably high agreement.
We also measure the reliability of the agreement with Fleiss' kappa score (0.32 on average), which represents fair reliability for human rating~\cite{FQ15}.
We finally annotate the image-based and fused-based clusters with KeyBert (ngram) because of the reasonable length of phrases and stable extraction quality.
From the three candidates provided by KeyBert (ngram), we then identify the POS tags for each token and select the phrase with at least an adjective, followed by one or more nouns (this allows us to generate meaningful annotation phrases).
If there is no such candidate, we return the top-1 candidate as the annotation.

\mypara{Results}
We randomly present one of the extracted key phrases with the Hillary cluster as an example.
The extracted phrases capture the keyword ``hillary'' which reasonably represents the entire cluster.
Also, the adjective or noun before/after ``hillary'' captures the contextual information, e.g., ``president-hillary-running'' in \autoref{figure:cluster_sum}.
We also visualize the top-100 clusters' annotations in \autoref{figure:annotation} by projecting the centroid embedding to a 2-dimensional space using t-SNE~\cite{MH08}.
Theoretically, meme clusters that are highly related in semantics stay closer in the high-dimensional space, and so is their 2D space after the projection.
Image-based and fused-based clustering results have commonalities in the embedding projection, e.g., recognizing the communities of politicians related in the US election and Praise Kek related communities, which are marked with green boxes.
However, we also highlight disparities between the two clustering results with red boxes.
While the fused-based clustering (\autoref{figure:annotation_sum}) identifies the cluster ``nazi-pepe'' as a member of the Hitler related theme, in the image-based clustering, ``nazi-frog'' cluster in \autoref{figure:annotation_image} is projected in the Praise Kek area as it weights the major visual feature (frog) far more than Nazi symbolism.
Another typical example is the cluster of Justin Trudeau, annotated as ``government-trudeau'' and ``trudeau-prime'' in two clustering annotations.
Image-based clustering places the cluster near other politicians, such as ``delegates-ted'' and ``rally-bernie,'' meanwhile, fused-based clustering manages to put it near ``flag-canada'' and ``leafposting-canadians,'' despite sizeable visual differences.
These observations validate the effectiveness of clustering, especially multimodal clustering and automatic annotation.

\begin{table*}[!t]
\centering
\caption{The 35 identified communities ranked by hate score. 
We report the name, the number and percentage of included clusters, the percentage of included posts, the hate score, and the centroid cluster.} 
\label{table:communities_info}
\setlength{\tabcolsep}{0.1em}
\scalebox{0.75}{
\begin{tabular}{lcccl|lcccl}
\toprule
Communities &  Clusters (\%) & Posts (\%) & Hate Score & Centroid Cluster  & Communities &  Clusters (\%)& Posts (\%)& Hate Score & Centroid Cluster  \\
\midrule
Holocaust            &48 (3.9\%) &3.1\%	&0.54  &holocaust-Jews         &Nazi Pepe          &43 (3.5\%) &3.3\%  &0.28  &warlord-gangs \\ 
Jews Posts           &58 (4.7\%) &3.9\% &0.49  &merchant-jew           &Race \& Society    &60 (4.9\%) &11.4\% &0.28  &fornication-data \\
Jews \& Minority     &20 (1.6\%) &1.4\%	&0.48  &largest-pizzagate      &Reddit-Plebbit	   &28 (1.8\%) &0.9\%  &0.28  &mfw-australians \\ 
African              &16 (1.3\%) &1.0\%	&0.47  &zimbabwe-movement      &Lybia	           &26 (2.1\%) &1.3\%  &0.27  &sowell-gaddafi \\
Illegal Immigration  &30 (2.4\%) &1.8\%	&0.45  &niggers-lmao           &Comics Like		   &21 (1.7\%) &1.1\%  &0.27  &drawing-pepes  \\ 
Refugees in EU       &21 (1.7\%) &1.1\%	&0.42  &germans-refugees       &American Posts	   &34 (2.8\%) &2.3\%  &0.26  &usa-flag \\
Jews Religion        &44 (3.6\%) &2.8\% &0.38  &highlanders-regiment   &Canada	           &23 (1.9\%) &2.3\%  &0.26  &canadabros-shitpost \\
Adolf Hitler         &72 (5.9\%) &7.2\% &0.38  &hitler-mein            &European Politics  &46 (3.7\%) &3.7\%  &0.26  &british-truly \\ 
Muslim               &25 (2.0\%) &1.5\% &0.37  &arabs-mfw              &White Supremacists &76 (6.2\%) &5.3\%  &0.24  &molyneux-stefan \\
Memeball (Meme)      &35 (2.8\%) &1.9\% &0.34  &memeballs-memeball     &Pepe \& Kek	       &45 (3.7\%) &2.6\%  &0.24  &chaos-pepe \\
Chinese \& Communism &20 (1.6\%) &1.4\%	&0.33  &socialism-gommunism    &Donald Trump	   &85 (6.9\%) &5.6\%  &0.23  &foods-trump \\
Worship Kek	         &46 (3.7\%) &2.8\% &0.31  &fat-boogie             &Thoth \&Skeleton   &40 (3.3\%) &2.2\%  &0.21  &frogs-pepe \\
Australian		     &22 (1.8\%) &1.5\%	&0.31  &political-memeball     &White Nationalists &31 (2.5\%) &1.8\%  &0.20  &communist-propaganda \\
Jews \& Talmud       &52 (4.2\%) &9.3\% &0.31  &page-talmud            &White Supremacy	   &46 (3.7\%) &3.2\%  &0.19  &worship-chaos \\
Spurdo (Meme)        &24 (2.0\%) &1.5\%	&0.30  &example-sage           &Politicians		   &47 (3.8\%) &7.1\%  &0.17  &clinton-health \\
Russian		         &17 (1.4\%) &1.5\%	&0.29  &blacks-smarter         &Others	           &34 (2.8\%) &1.8\%  &- 	  &- \\
\bottomrule
\end{tabular}
}
\end{table*}

% ----------------------------------------------------
\subsection{Hate Analysis}
\label{subsection: hate_analysis}
% ----------------------------------------------------

Online users generally initiate and spread hateful content through meme images and textual opinions.
Many hateful speech detection services are developed to automatically score the input text in terms of toxicity, abuse, etc., such as Google's Perspective API~\cite{Perspective}, Rewire~\cite{rewire}, and toxic Bert~\cite{Detoxify}.
Meanwhile, highly toxic meme images are studied insufficiently due to the absence of a large labeled toxic dataset.
We now conduct a hate assessment on the meme cluster basis using the bridge that CLIP has built between memes and textual comments.
By doing this, we help the platform moderators to find out: 1) which groups are the hateful targets of 4chan users' views? and 2) with what memes are they spreading hateful sentiments? 

Before conducting the hate analysis, it is essential to clarify the concept of Hate studied in our research.
We align the Hate definition with that of the United Nations~\cite{unitednations}, and summarize it as ``speech, writing, or behavior that attacks a person or a group based on one's identity.''
We refer to hate based on one's identity as ``Identity attack.''
Meanwhile, we exclude abusive language against a specific person, e.g., ``I hate you.''
The reason we apply the definition is twofold.
First, 4chan's /pol/ is filled with toxic, abusive, and insulting phrases, e.g., ``fucking,'' ``damn,'' and ``stupid.''
The occurrence of these words will make the majority of the sentences immediately hateful according to the other general definition~\cite{KFMGSRT20}.
Here, we focus on Identity attack instead of these toxic ``noises.''
Also, identity attack topics are prevalent in 4chan's /pol/, especially antisemitism and islamophobia as studied in~\cite{GZ22,ZFBB20}.

\mypara{Hate measurement}
We measure the hate score of texts using Google's Perspective API~\cite{Perspective} and Rewire~\cite{rewire}.
Perspective API uses machine learning models to identify abusive comments on different dimensions like Toxicity, Insult, Profanity, Identity attack, Threat, etc.
Here we use the Identity attack score as our hate indicator to reflect on the online hatred targeting a group of people based on their identity.
Rewire is another tool for detecting hate speech targeting identities.
For each text, it returns the predicted label (``hateful,'' ``non-hateful'') with a confidence score.

We conduct the hate assessment on the fused-based clustering results as they combine both text and images.
We obtain 1,229 clusters after filtering out the clusters with less than 30 samples in 17,654 clusters.
To measure the hate score of each cluster, we extract all the textual posts within the same cluster and obtain both the Identity attack score returned by the Perspective API and the Hateful label returned by the Rewire.
For the Perspective API, the text is considered hateful if the returned confidence score is larger than 0.7, according to \cite{perspective_score}.
A textual post is believed to be hateful if at least one of the above APIs returns a hateful label.
We calculate the fraction of hateful textual posts in all posts of a cluster as the Hate score, which indicates the level of users attacking the person or group based on their identity.
The rationale for transferring the hate presented by texts to meme clusters is that the fused meme embeddings contain textual information.
Eventually, 1,229 memes clusters are measured in terms of Hate.
These clusters contain 93,501 posts and account for 9.4\% in our selected dataset (note that the percentage is small due to the noise level of the clustering algorithm and the fact that we remove clusters with less than 30 samples).

\begin{figure*}[ht]
\centering
\begin{subfigure}{0.65\columnwidth}
\includegraphics[width=\columnwidth]{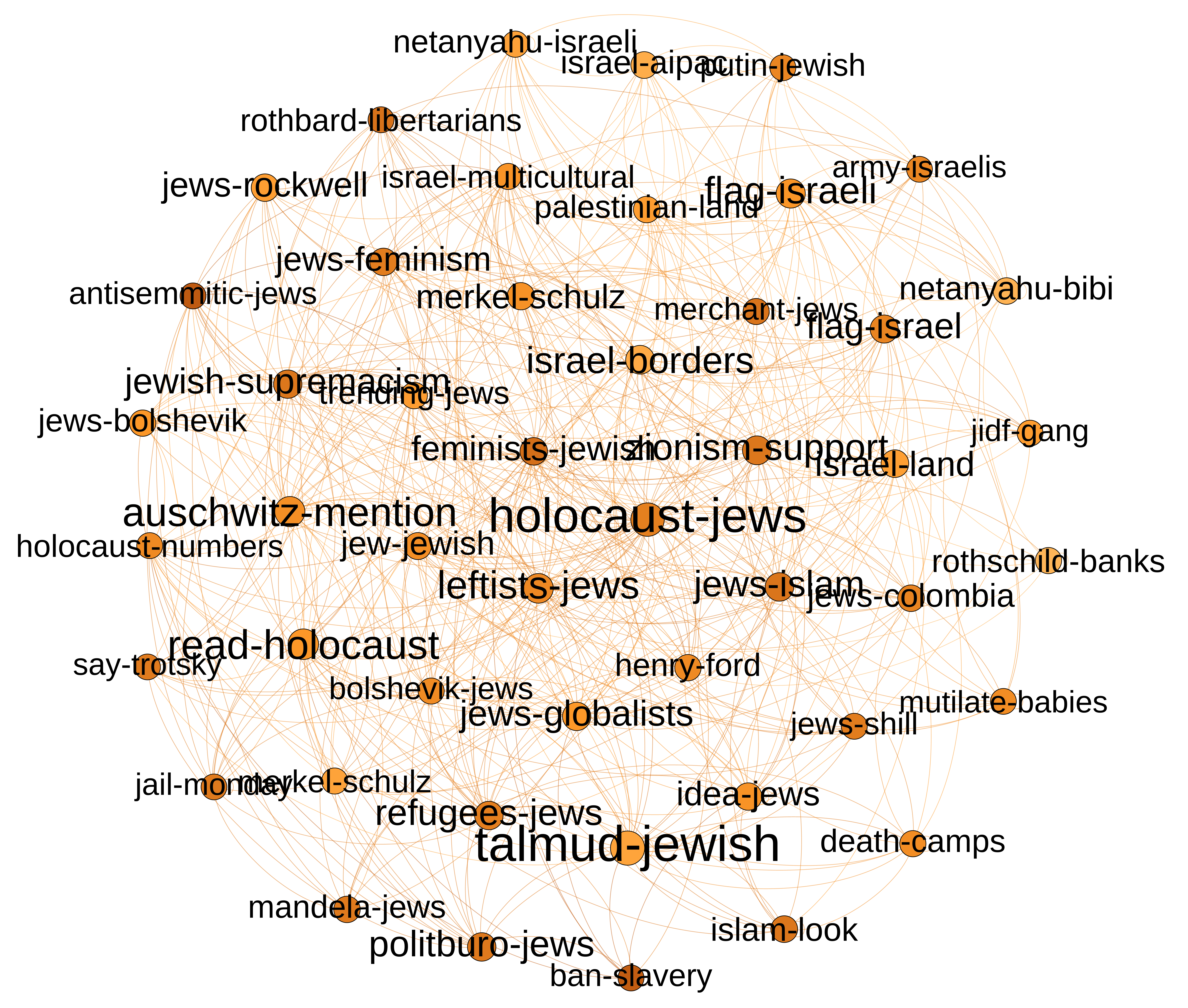}
\caption{Holocaust}
\label{figure:holocaust}
\end{subfigure}
\hfill
\begin{subfigure}{0.65\columnwidth}
\includegraphics[width=\columnwidth]{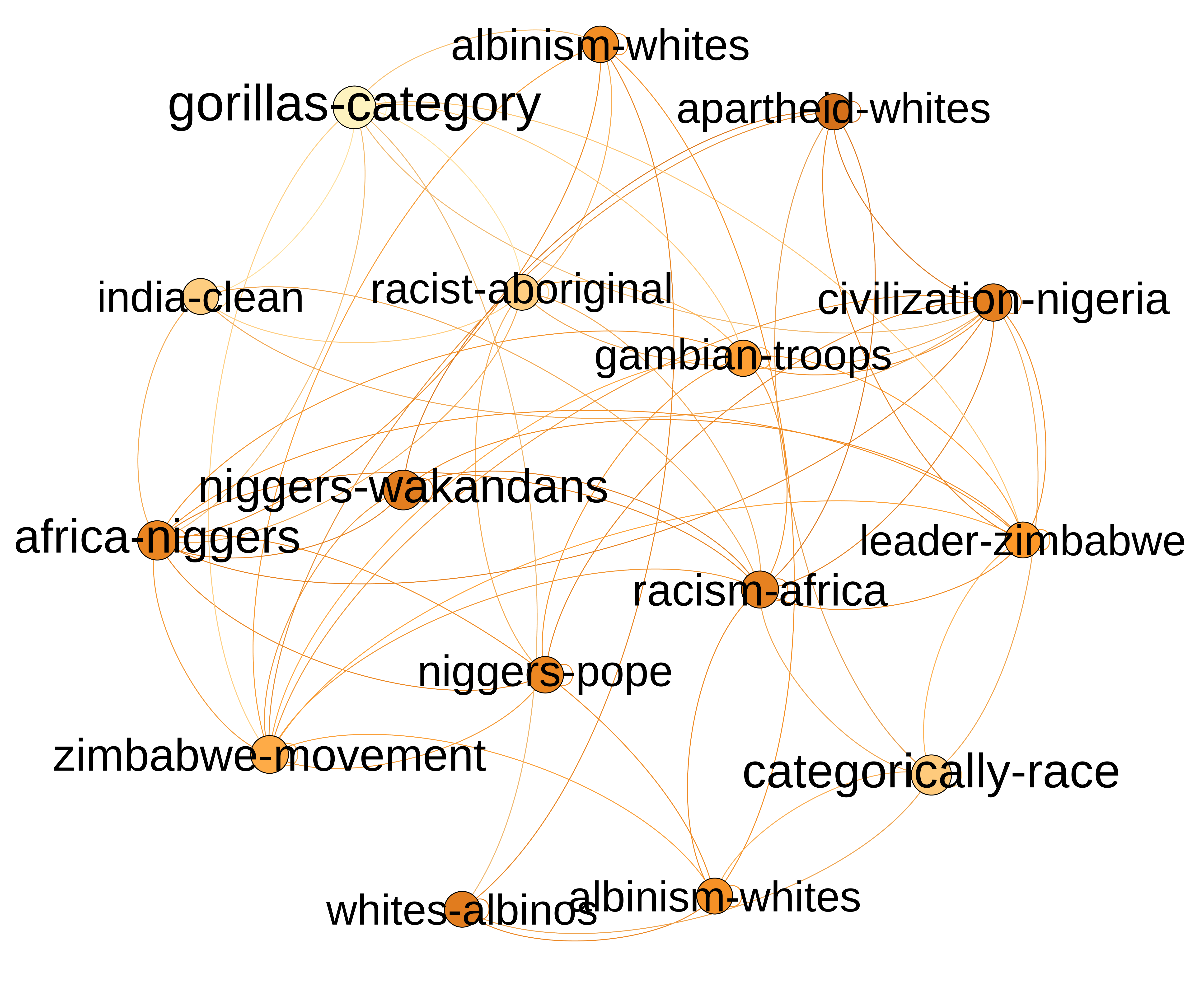}
\caption{African}
\label{figure:african}
\end{subfigure}
\hfill
\begin{subfigure}{0.65\columnwidth}
\includegraphics[width=\columnwidth]{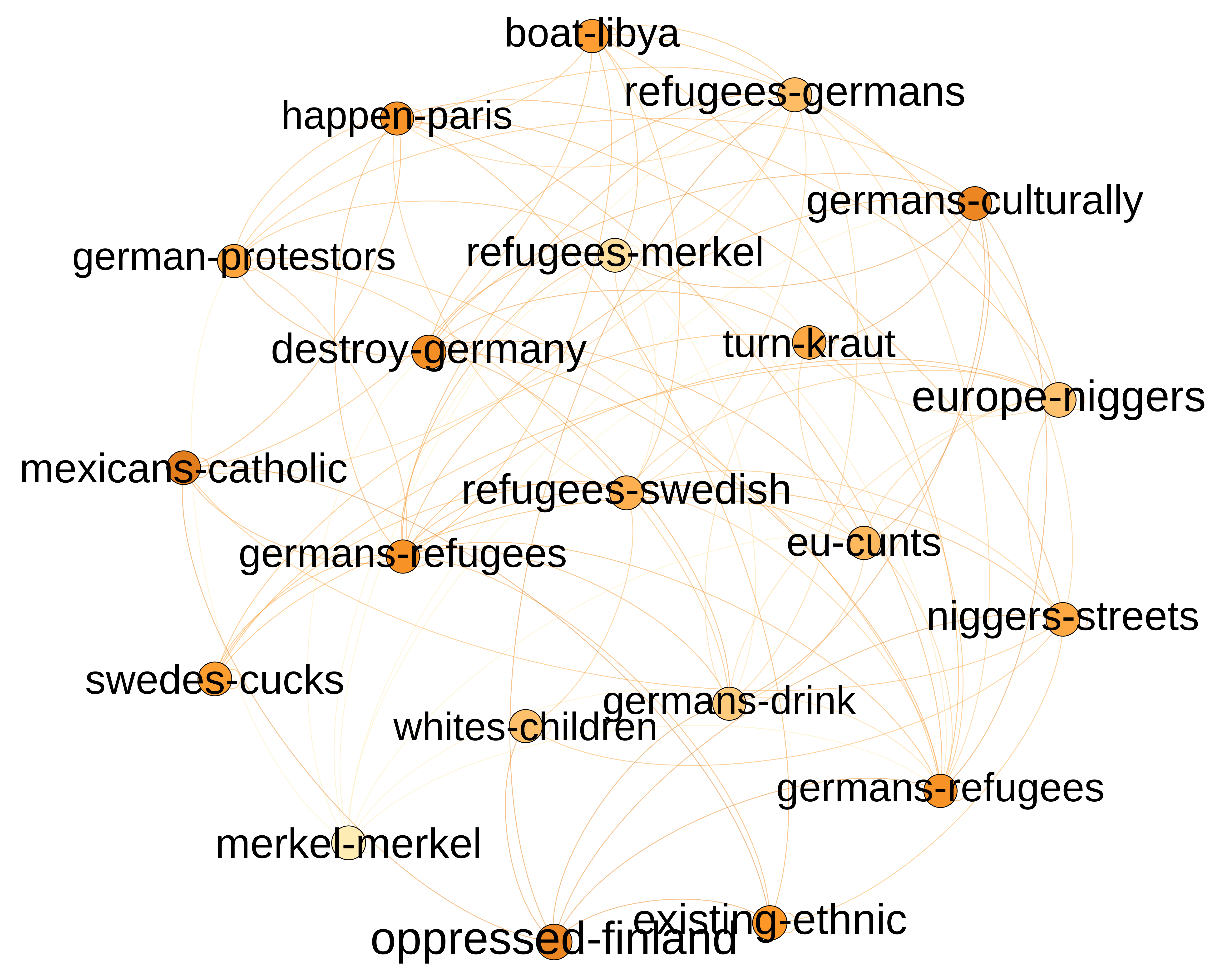}
\caption{Refugees in EU}
 \label{figure:refugees}
\end{subfigure}
\caption{Visualization of three communities with high hate scores. Each node represents a cluster. 
We distinguish the hateful level of clusters with color; deeper color corresponds to a higher hate score.}
\label{figure:3_communities}
\end{figure*}

\mypara{Community detection}
By examining the most hateful clusters, one might conclude the people/groups that are primarily resented in the view of 4chan users.
To further reduce the complexity of understanding all 1,229 clusters and primary hate targets, we construct the cluster graph and perform community detection to reduce thousands of clusters to dozens of communities.
The clusters are nodes $V$ in the graph $G$, and the semantic distance among clusters can be denoted as the weights of edges $E$.
We leverage the cosine similarity of two centroid embeddings of clusters to represent semantic distance.
To avoid excessive edges, we remove all the edges whose weights are less than the 98 percentile of the edge weights.
Community detection is a technique that reveals the hidden relation of nodes in a graph and identifies densely connected nodes with commonalities.
We employ the Louvain method~\cite{MFFP11} to identify the communities.
The goal of the algorithm is to maximize the modularity~\cite{N06} of the communities, where the modularity measures the ratio of the high density of edges inside communities to edges outside communities.
The value of modularity generally falls between -0.5 and 1, indicating the increasingly better modular partition.
We identified 35 communities in \autoref{table:communities_info} based on fused embedding clusters, and the mean modularity value is optimized to 0.39, which indicates a fair partition\cite{MFFP11}.
We measure the Hate score of a community by calculating the fraction of hateful posts in all posts in this community (e.g., if the fraction is 0.05, it means 5\% of all posts in the community are considered hateful).
Communities include various meme clusters, and we name each community based on its theme by looking into the automatic cluster annotations.

\mypara{Findings}
We make several observations based on the hate scores of 35 communities.
First, the Jewish community has become the most prevalent target of hateful memes on 4chan.
Many communities are antisemitism-related with high hate scores, e.g., Holocaust, Jews Posts, Jews\&Minority, Jews Religion, and Jews Talmud.
We demonstrate the details of the Holocaust community in \autoref{figure:holocaust}, where 4chan users spread hate on the discussion of merchant-Jews, Jews-globalist, Israel issues, etc.
This indicates that the above meme clusters usually incite hateful sentiments against Jewish and require moderators' intervention.
Second, Africans are also a severe hate target in 4chan but with fewer clusters than antisemitism.
\autoref{figure:african} displays the major topics regarding Africa, from which we observe people pour hate related to Gambia, Zimbabwe, and Nigeria.
Third, immigrants and refugees are also vulnerable groups that 4chan users disrespect.
The detailed clusters in Refugees in EU in \autoref{figure:refugees} imply that people show negative attitudes towards refugees in Europe, especially refugees in Germany.
In addition, Muslims, Chinese, and Australians are often the hate targets for spreading hateful memes based on \autoref{table:communities_info}.

% ----------------------------------------------------
\section{Hateful Memes Evolution}
\label{section: evolution}
% ----------------------------------------------------

In this section, we use semantic regularities in CLIP's embeddings to understand the evolution of hateful memes.
Here we use all 12.5M image-text pairs in our 4chan dataset.

For memes serving as a hateful signal, e.g., the Happy Merchant meme, users tend to express their negative feelings by combining the hateful signal with other elements like persons, countries, and organizations.
The resulting product is referred to as a \emph{Variant}, and the element used for creating the variant is named \emph{Influencer}.
For instance, the Trump version of Happy Merchant is an example of a variant, with the image showing Donald Trump serving as the influencer (see \autoref{figure:manipulations}).
Also, as can be seen in \autoref{figure:manipulations}, influencers can be either image influencers or textual influencers.
We study variants and influencers by extracting semantic regularities.
In \autoref{subsection: visual_semantic_regularities}, we demonstrate how to identify variants globally in the dataset and estimate the most likely image influencer with the case study of Happy Merchant.
We also identify hateful variants in a directed manner by pre-selecting the textual influencers in \autoref{subsection: visual-linguistic_semantic_regularities}.
We further study the temporal dynamics of identified variants of Happy Merchant in \autoref{subsection: visual-linguistic_semantic_regularities}.
Additionally, to demonstrate the generalizability of our framework, we present a similar analysis for the Pepe the Frog meme in \autoref{appendix: case_study_pepe} in the Appendix.

% ----------------------------------------------------
\subsection{Visual Semantic Regularities}
\label{subsection: visual_semantic_regularities}
% ----------------------------------------------------

\begin{figure}[!t]
\centering
\includegraphics[width=\columnwidth]{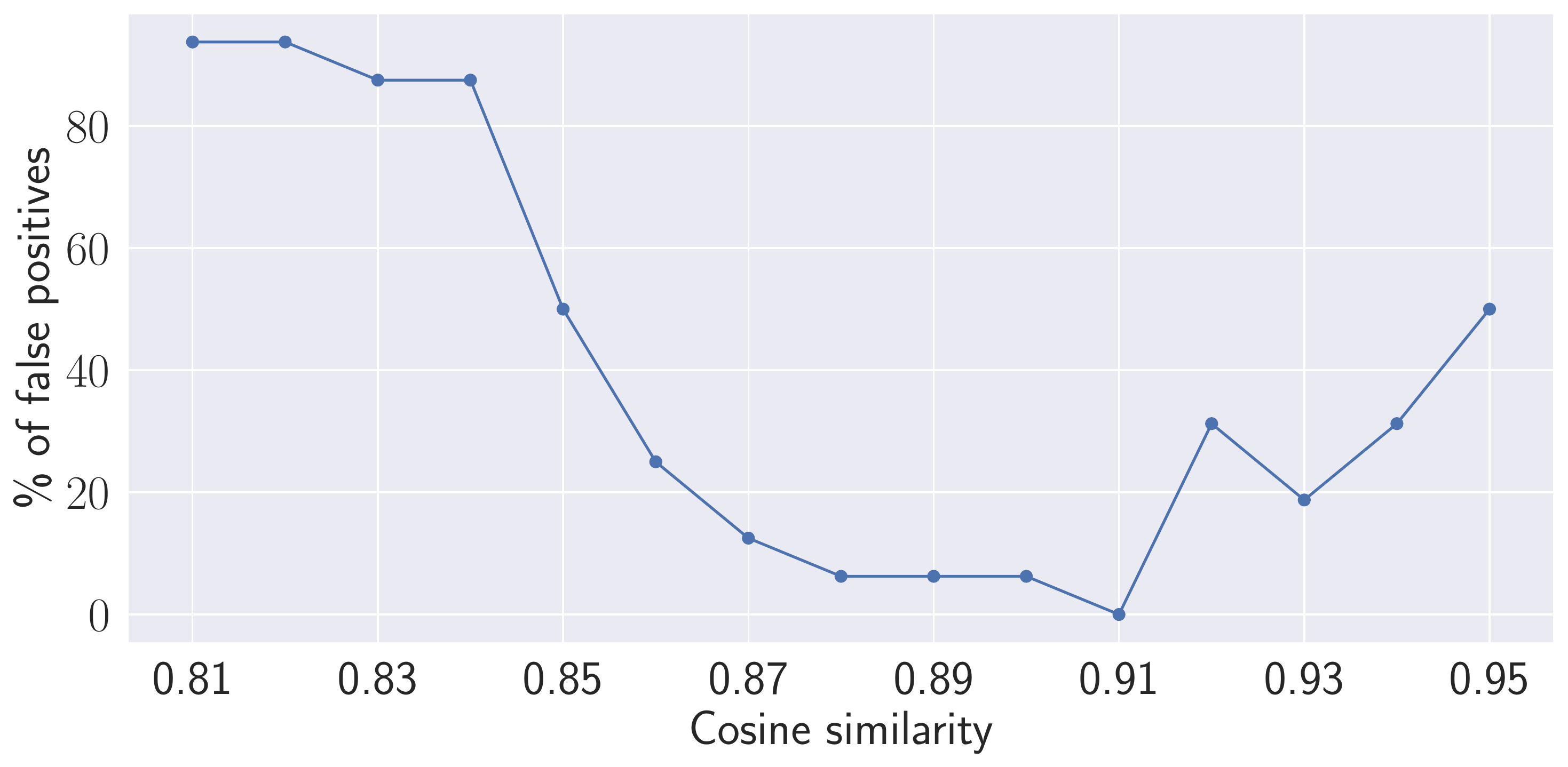}
\caption{Percentage of false positives for varying cosine similarity threshold for the Happy Merchant meme.} 
\label{figure:false_positives}
\end{figure}

\begin{figure*}[!t]
\centering
\includegraphics[width=\textwidth]{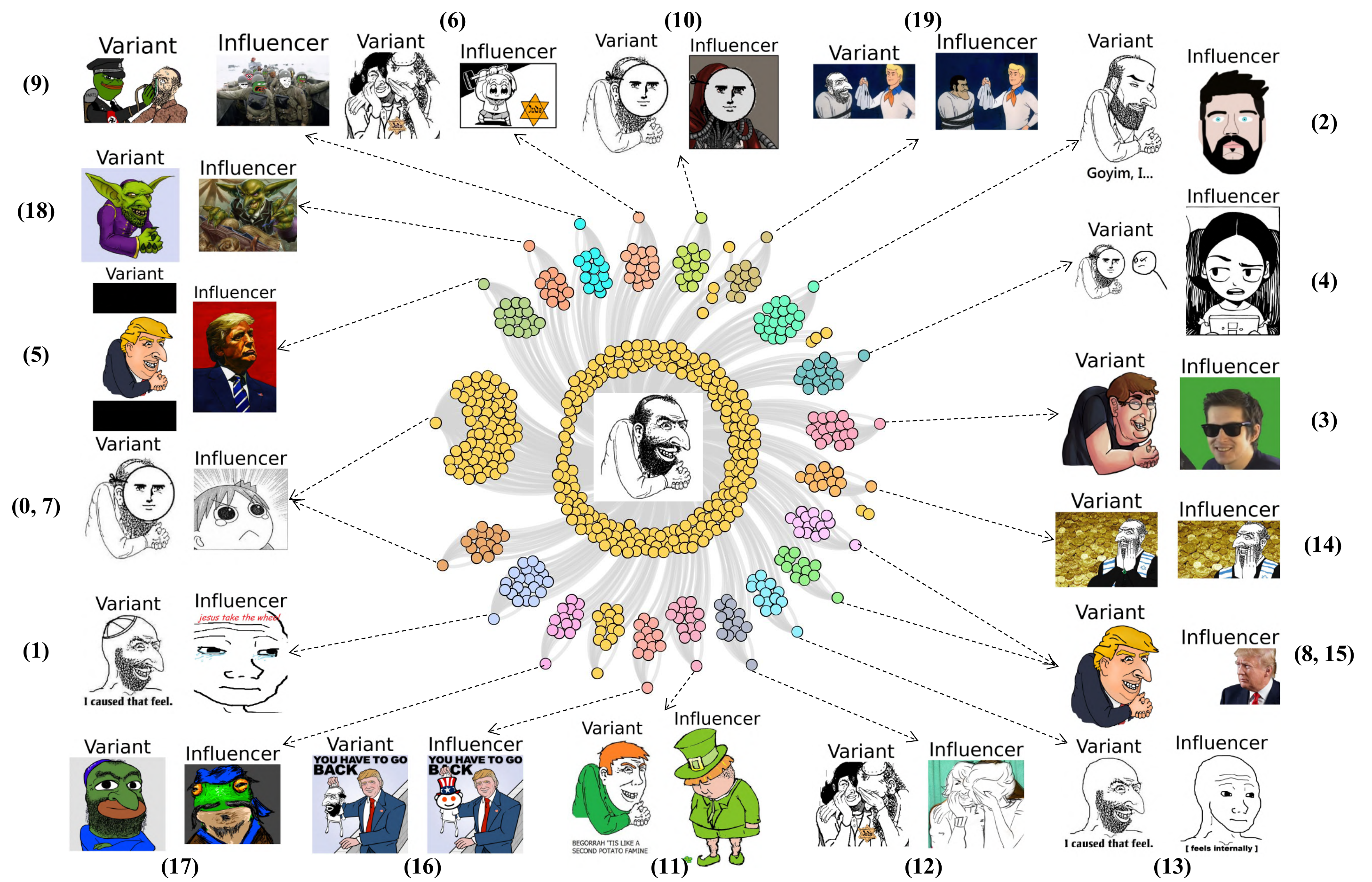}
\caption{The top-20 communities in the ecosystem of Happy Merchant. Colors differentiate the communities, and we annotate each community with two images: one of the variants on the left and its potential influencer on the right. 
We also include the community index to assist us in referencing the communities in the main text.} 
\label{figure:evolution_image_image}
\end{figure*}

\mypara{Pipeline}
This method aims to locate the variants and identify the corresponding influencers via the operations on image embeddings.
The intuition of finding variants is that the variants are partially similar but not identical to the original hateful image ($m_o$) because these variants share visual features and semantics with the original $m_o$.
Concretely, with the popular hateful image fixed, we first manually determine a lower bound ($t^v_{lower}$) and upper bound ($t^v_{upper}$) of cosine similarity where all the images in the dataset are considered as variants if their embedding similarities with $m_o$ are within this range.
For each meme variant $m_v$, we retrieve the candidates potentially serving as image influencers $m_i$.
By calculating the top-$k$ cosine similarities $cos(e_o+e_c, e_v)$ where $e_c$ is every candidate image in the dataset, we obtain top-$k$ influencer candidates.
Generally, we directly take the top-1 image as the influencer.
During the retrieval process, we observe that when the image is highly similar to the meme variant $m_v$, the resulting embedding similarity is prone to be extremely high; thus, these unexpected images also get into the set of top-$k$ influencer candidates.
To alleviate this issue, one could mask off the highly similar images before selecting top-$k$ influencer candidates by setting another threshold ($t^i_{upper}$).
Finally, with the retrieved triplet $(m_o, m_v, m_i)$, we record the cosine similarity and discard the triplet if the similarity is below the threshold ($t^i_{lower}$).
Note that the selection of these thresholds depends on the use case; hence we recommend trying various thresholds and assessing a sample of the results manually.
In our experiments, we find that determining proper thresholds takes approximately 1-2 hours of manual work.
In the future, we plan to develop solutions to automatically identify proper threshold values to reduce the required load for content moderators.

\mypara{Case study}
Happy Merchant is one of the most prevalent images for spreading antisemitic ideologies~\cite{ZCBCSSS18}.
It is often blended with other elements and produces new variants to transfer the hate targets.
Here, we adopt Happy Merchant as the original image meme and apply the framework introduced above to study its evolution.
Applying the above pipeline, we first identify the variants of Happy Merchant by setting [0.85, 0.91] as the similarity range ($t^v_{lower}, t^v_{upper}$).
Specifically, we increase the similarity threshold from 0.81 to 0.95 with the step of 0.01, where at each threshold, we randomly select 16 images whose embeddings have the same similarity as the threshold and manually judge if they are all blended products derived from the original Happy Merchant.
As the threshold increases, the searched images tend to present more apparent visual features than the original image.
As \autoref{figure:false_positives} shows, if the lower bound threshold is smaller than 0.85, we observe that a high percentage of the searched images are false positives (unrelated images).
Similarly, if the higher bound threshold is greater than 0.91, the false positives will also rise as more images are visually identical to the original Happy Merchant meme instead of its variants.
Empirically, considering that we intend to identify the variant-influencer pairs simultaneously, we suggest adopting a relatively smaller lower bound to include as many variants as possible, then filter out the pairs with inaccurate influencers at a later stage.
Note that these thresholds depend on the original image, and if we study the evolution of other memes, e.g., Pepe the Frog, new thresholds are required.

When identifying image influencers, we first set identical $t^v_{upper}$ and $t^i_{upper}$ as 0.91 since both exclude highly similar images.
We then exclude the variant-influencer pairs if the cosine similarity ($cos(e_o+e_i, e_v)$) of the variant embedding and the summed embedding is lower than 0.94 (we set the threshold following the same methodology as 
the one used for the identification of the variants).

We manage to identify 3,321 pairs of variants and the top-1 influencers.
To evaluate the accuracy of identifying variants and influencers, we manually annotate 100 randomly selected image pairs.
The annotation is conducted by the three authors of this paper independently.
We take the majority agreement as the final annotation and report a 3-person-agreement score of 0.51 and 2-person-agreement score of 0.66.
Based on our annotations, 78\% of the variants and 53\% influencers have been successfully identified.
We find 22\% of the identified variants are false positives.
By inspecting the false positives, we find that some pencil-sketched memes are misclassified as Happy Merchant variants, likely due to a similar drawing style.
Images of classic Jewish people and Adolf Hitler are also often falsely grouped into variants because of their close semantic distances.
Content moderators can adopt a more conservative strategy by increasing the $t^v_{lower}$ threshold to reduce the number of false positives (and inevitably missing some variants).
We recommend the moderators select the thresholds that fit their moderation strategy.
Also, by looking into the top-10 images instead of top-1 for identifying influencers, and repeating our annotations, we find that the identification rate increases to 61\%, highlighting that moderators can potentially review the top-$N$ images to identify influencers.

\begin{figure*}
\centering
\includegraphics[width=\textwidth]{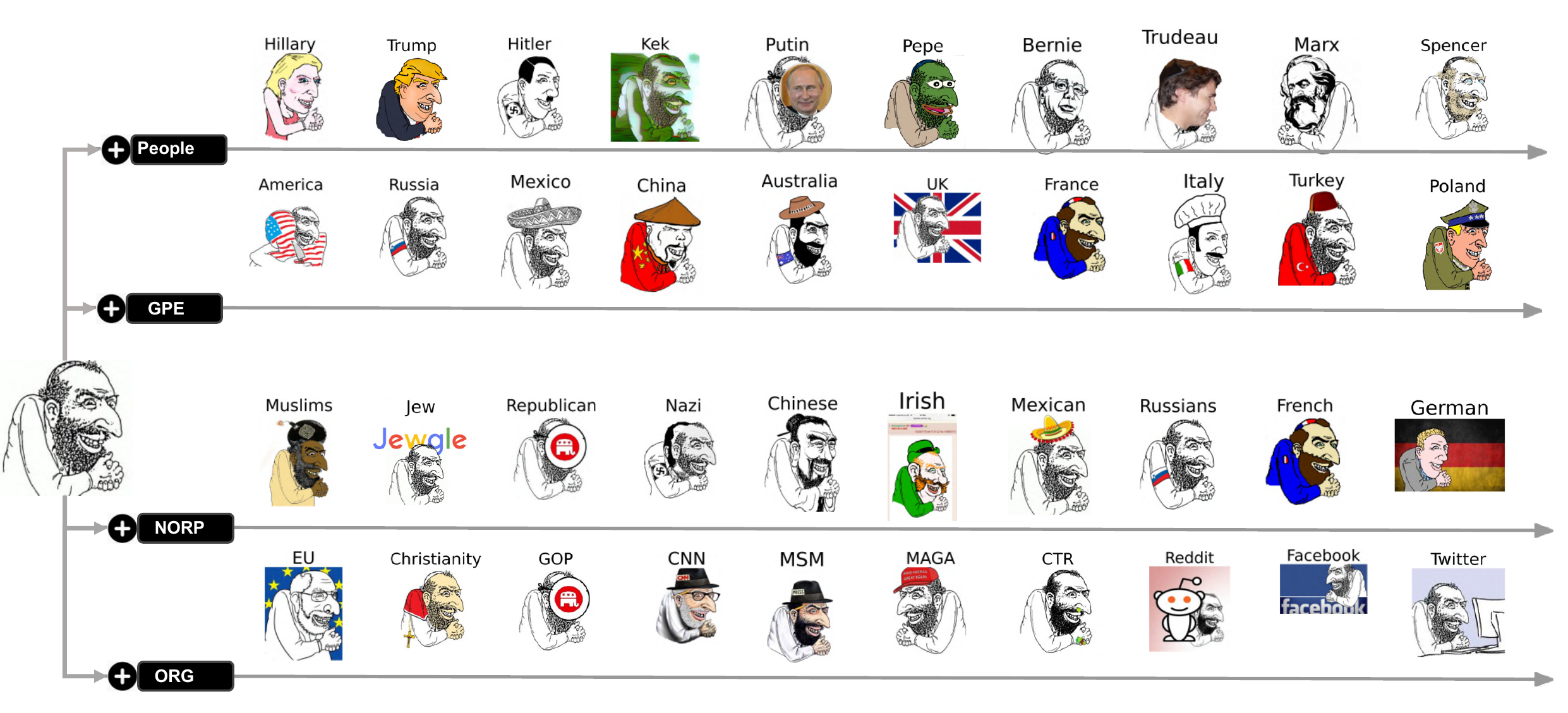}
\caption{Happy Merchant variants influenced by the four types of entities (textual influencers, 10 examples for each type). We show all these different hateful meme variants to demonstrate the extent and true nature of the hateful content problem online and raise awareness about these symbols and their variants.}
\label{figure:evolution_image_text}
\end{figure*}

To further probe the prevalence of different variants, we build an undirected graph with the retrieved data where each image is a node and the edge denotes the summation relation between images, e.g., a variant connects the original image and its influencer.
The graph contains 5,279 nodes (images) and 6,656 edges (variant-origin, variant-influencer).
For better visualization, we perform community detection and select the top-20 communities (see \autoref{figure:evolution_image_image}).
We mark the communities with both colors and numbers (community ids) beside the images.
A smaller community id indicates a larger number of members in this community.
One interesting observation is that, for most of the communities, there is usually one influencer node shared by multiple variant nodes.
This indicates that there are many visually identical variants that are influenced by the same image.
To inspect what the common influencers are and how the variants are influenced, we annotate each community with two images: the variant and the influencer.
In detail, we directly visualize the node with the largest degree in each community as the influencer and visualize a random node that each influencer connects in the community as the variant.

\mypara{Findings}
Based on the top-20 communities of variants in the evolution of Happy Merchant, we find that the most popular variant of Happy Merchant is in communities 0, 4, 7, and 10, where the merchant wears a mask of a lovely face.
This might indicate that 4chan haters advocate that Jewish are hypocritical and good at disguising.
The influencers in the above communities might change; however, they all possess the characteristics of ``friendliness'' and ``innocence.''
Also, Happy Merchant is prone to combine with real persons, such as Trump in communities 5, 8, and 15, and other persons in community 3, to reflect 4chan users' opinions on the real person.
Happy Merchant is often fused with other classic memes, one of which is the Feels Guy meme~\cite{feelsguy} in communities 1 and 13.
Another one observed is Pepe the Frog~\cite{pepethefrog}, shown in communities 9 and 17.
Additionally, the variants can be developed by combining multiple elements with Happy Merchant, e.g., community 9 indicates both the frog and Nazi ideology influence the variant, and the variant in community 16 is influenced by both Donald Trump and the Reddit Meme~\cite{reddit}.

% ----------------------------------------------------
\subsection{Visual-linguistic Semantic Regularities}
\label{subsection: visual-linguistic_semantic_regularities}
% ----------------------------------------------------

\mypara{Pipeline}
Unlike retrieving variants and image influencers only via image embedding operations (i.e., semantic regularities on image embeddings), we can also discover new variants by pre-defining a set of textual influencers and identify images that are generated because of the fusion of an image and a specific textual influencer (i.e., Semantic regularities on image and text embeddings).
Given that hateful content online is usually influenced by real-world events that usually involve various entities like persons, countries, or organizations, in this work, we define a set of textual influencers by leveraging techniques from Natural Language Processing, particularly, Named Entity Recognition.
Concretely, we extract named entities from the posts on 4chan /pol/ dataset using the spaCy~\cite{spacy2} python library.
Named entities are divided into different categories, such as People (e.g., Donald Trump), Geo-Political Entities (GPE, e.g., America), Nationalities, Religious or Political Entities (NORP, e.g., Muslims), Organizations (ORG, e.g., EU), numbers, and dates.
In this work, we select the top-30 most frequent entities from the following categories: People, GPE, NORP, and ORG, as these categories are related to the topics discussed in /pol/ and are likely to influence the creation of hateful memes.
By having a fixed original hateful image, we compute the fused embedding ($e_f$) by performing a weighted summation on the original image embedding ($e_o$) and textual embeddings ($t_i$) of the selected entities, such as $e_f = 0.2*e_o + 0.8*t_i$.
The weight selection is explained in \autoref{subsection: flexible applications}.
We then retrieve the top-$k$ most similar images ($m_v$) as the variants by computing $cos(e_v, e_f)$.
Using this approach, we aim to perform a targeted identification of hateful variants by using a set of pre-defined textual influencers extracted from named entity recognition.
For instance, by fixing the original image to the Happy Merchant Meme and the textual influencer to ``Donald Trump,'' we can identify Donald Trump's Happy Merchant variant in an automated and systematic manner.

\mypara{Case study}
We apply the above-mentioned pipeline to identify variants of the Happy Merchant meme in a directed manner.
As mentioned, we select the top-30 entities as the textual influencers from 4 categories: People, GPE, NORP, and ORG, and retrieve the top-$k$ closest image embeddings in the image embedding space.
Specifically, we extract the top-2 most similar images and select the one that is more popular on our dataset (in terms of the number of posts that appear in our dataset); we do this as we aim to identify popular variants and conduct temporal analysis later.
We further perform a manual inspection on all 116 identified variants (remove 4 noisy entities), from which we discover 75 variants that are successfully fused with Happy Merchant and entity semantics.
The annotation is conducted, again, by three authors of the paper independently.
We take the annotation of the major agreement as the final annotation, and we report that the 3-person-agreement is 0.59 and the 2-person-agreement is 0.63.
Specifically, there are 48.3\% entities in People, 76.7\% in GPE, 80.0\% in NORP, and 44.4\% in ORG that have the corresponding variants.
\autoref{figure:evolution_image_text} shows 10 variant examples for each category.
Overall, the retrieved variants preserve the structural feature of Happy Merchant and also the characteristics guided by the textual influencers.
For entities in GPE and NORP, e.g., country names, we observe a large possibility of these entities combining with Happy Merchant than other categories.
Furthermore, when Happy Merchant is fused with entities such as countries in GPE and nationalities in NORP, not only does the merchant's face adapt to the new nations, but the national flag is also often used to ``decorate'' the merchant or serves as a background.
For People, politicians are vulnerable to fusing with Happy Merchant since we find Happy Merchant variants for Hillary Clinton, Donald Trump, Bernie Sanders, Vladimir Putin, and Justin Trudeau.
We also find successful fused examples with other memes like Pepe the Frog, which are consistent with the findings in \autoref{subsection: visual_semantic_regularities}.
Additionally, for ORG, Happy Merchant is also prone to meddle with social platforms such as Reddit, Facebook, and Twitter, mainstream media such as CNN and MSM, and religions such as Christianity.

\begin{figure*}[ht]
\centering
\begin{subfigure}{0.49\textwidth}
\includegraphics[width=1\linewidth]{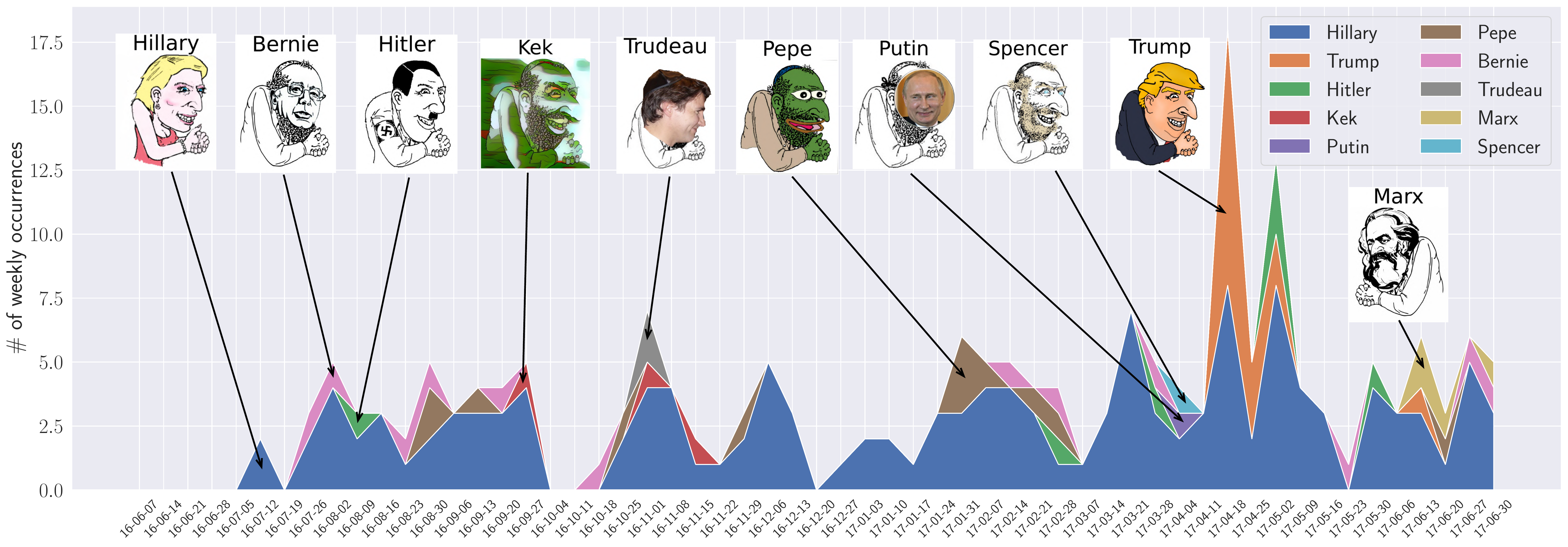}
\caption{People}
\label{figure:people} 
\end{subfigure}
\begin{subfigure}{0.49\textwidth}
\includegraphics[width=1\linewidth]{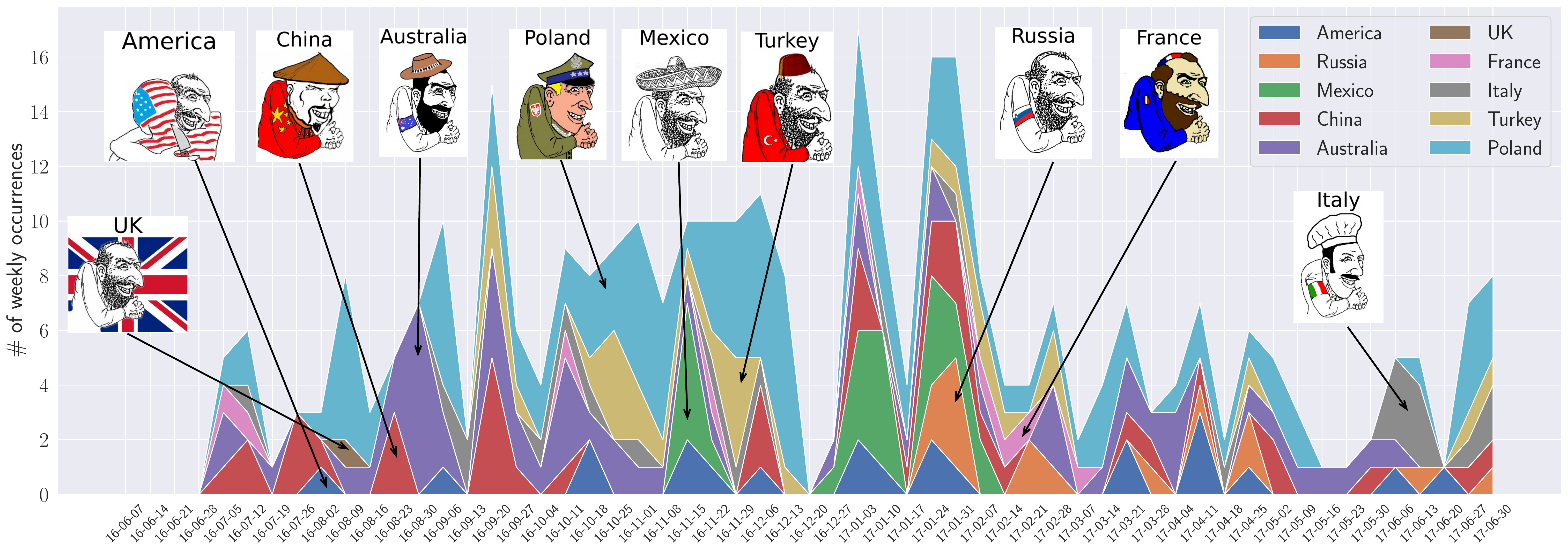}
\caption{GPE}
\label{figure:gpe}
\end{subfigure}
\medskip
\begin{subfigure}{0.49\textwidth}
\includegraphics[width=1\linewidth]{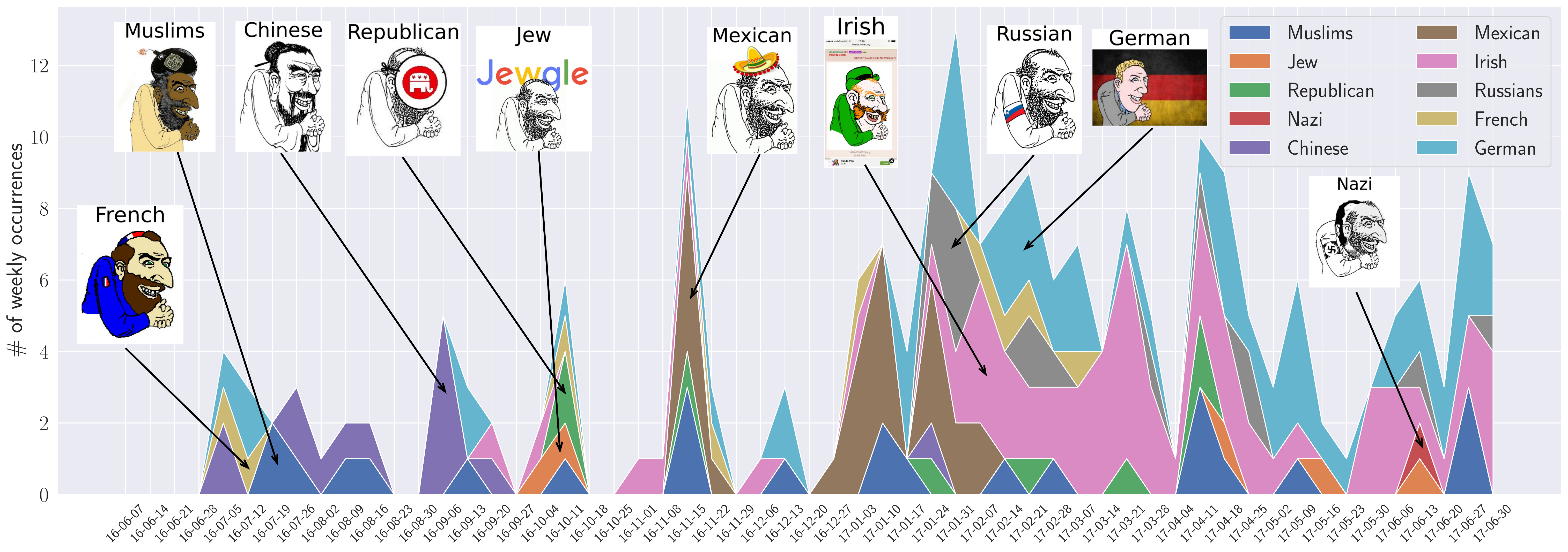}
\caption{NORP}
\label{figure:norp} 
\end{subfigure}
\begin{subfigure}{0.49\textwidth}
\includegraphics[width=1\linewidth]{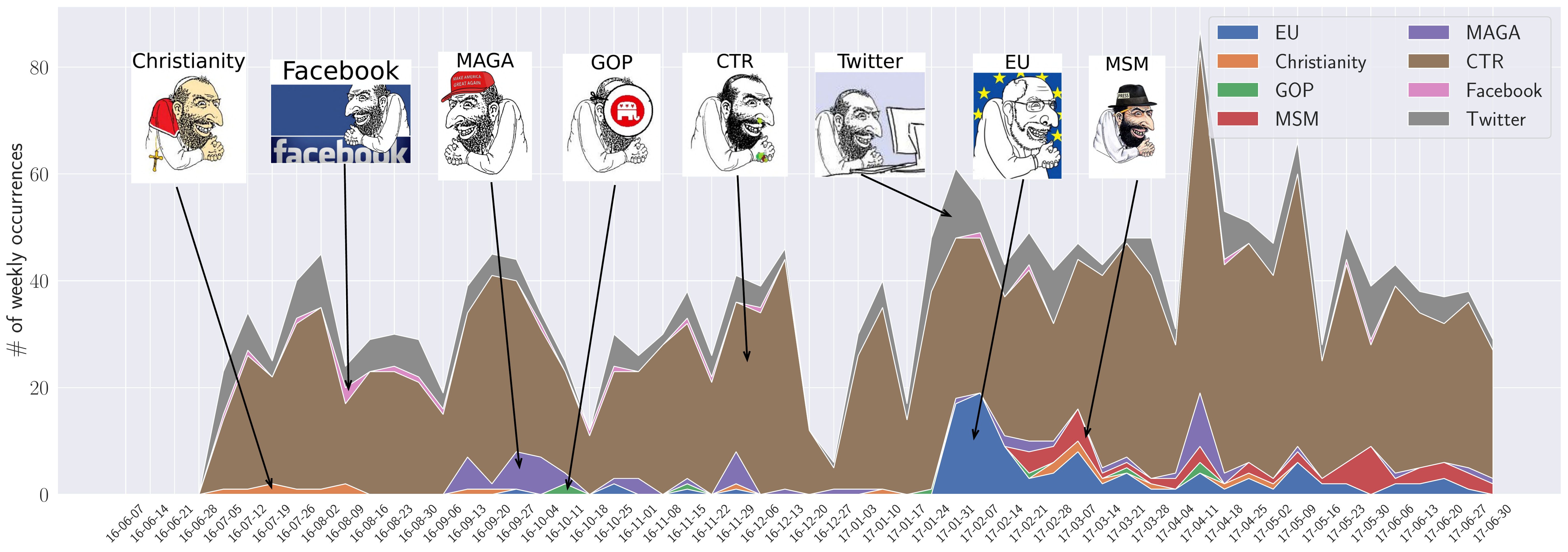}
\caption{ORG}
\label{figure:org}
\end{subfigure}
\caption{Number of posts including Happy Merchant variants per week.}
\label{figure:temporal}
\end{figure*}

\mypara{Temporal analysis}
Considering that memes evolve over time, it is natural that some hateful variants appear at specific points in time.
To study this phenomenon, we undertake a temporal analysis of the identified Happy Merchant variants.
In particular, for each Happy Merchant variant, we calculate the number of posts that include each variant, on a weekly basis, between June 30, 2016, and June 31, 2017.
Due to the fact that the same image could have many duplicates of different transformations, e.g., cropping and saving format, we use perceptual hashing (phash)~\cite{S17} here to account for the duplicates.
We group images according to their phashes and we consider them as the same image for our temporal calculations.
\autoref{figure:temporal} shows the number of posts, including each variant that is identified when considering the four categories of named entities.
We observe that Happy Merchant variants have different temporal patterns.
For instance, by looking at the variants extracted from people (\autoref{figure:people}), we observe that the Hillary Clinton variant is popular and appears consistently in 4chan's /pol/ throughout the period of our study, likely because of 4chan's opposition to Hillary Clinton's presidential run in 2016.
On the other hand, we observe other variants that are more concentrated on specific time periods; the Justin Trudeau variant is shared mainly during October 2016, while the Donald Trump variant is shared during April 2017, confirming the results of previous work~\cite{ZFBB20}.
Looking at the variants extracted for the GPE, NORP, and ORG categories (\autoref{figure:gpe}, \autoref{figure:norp}, and \autoref{figure:org}, respectively), we observe that most variants are consistently disseminated over the course of the time in 4chan's /pol/, which further highlights the large variety of antisemitic hateful connotations disseminated on 4chan.
Overall, we argue that such temporal analyses are useful to identify specific campaigns aiming to share specific hateful meme variants.
That is, to identify sharp increases of specific variants on social media platforms that can inform social media operators to moderate and mitigate the effects of the spread of hateful memes.

% ----------------------------------------------------
\section{Related Work}
\label{section:related work}
% ----------------------------------------------------

\mypara{Understanding and annotating memes}
Meme understanding is an evolving process.
Before the emergence of deep learning techniques, research works primarily focused on meme-spreading activities such as the diffusion process.
Wang et al.~\cite{WFVM12} models the diffusion process of memes spreading as hashtags via the agent-based model to understand meme popularity, diversity, and lifetime.
In their following work~\cite{WMA14}, they investigate the predictability of successful memes using historical patterns.
Dubey et al.~\cite{DMCR18} propose a meme embedding construction for memes with text overlaid on them.
They combine visual features and textual features such that the meme representations contain rich semantic information from both the image and the embedded texts.
By clustering on the meme embeddings, they group together memes with the same visual structure (template meme).
Differently, Beskow et al.~\cite{BKC20} leverage graph learning to find meme ``families.''
They first propose Meme-Hunter to find images on the Internet and identity them as memes or non-memes and then classify and cluster memes on Twitter into groups.
They put a special focus on characterizing meme usage in political conversations.
Methods for meme annotation are limited, with previous work~\cite{ZCBCSSS18} often relying on websites, e.g., (Know Your Meme)~\cite{KYM} to label meme images.
By contrast, we extract visual features and contextual features with a state-of-the-art model and construct a new meme embedding space that contains multimodal semantics.
We also leverage the connection between images and texts to provide an automatic annotation for memes in a self-explained manner.

\mypara{Hateful content detection}
Hateful content often combines different modalities such as image and text~\cite{SAADMFHSNC22}.
Many research efforts focus on hate speech detection.
Zahrah et al.~\cite{ZNG22} examine how the posting behavior of hateful communities on Reddit and 4chan changed during the 2020 US election.
They employ NLP techniques such as topic modeling and sentiment analysis tools.
Zannettou et al.~\cite{ZFBB20} provide a quantitative approach to studying online antisemitism.
They study the antisemitic language by studying the dynamic distances between word embeddings over time.
Fatemeh et al.~\cite{TSLBSZZ21} and Shen et al.~\cite{SHBBZZ22} characterize the evolution of Sinophobic language after the COVID-19 outbreak.
Other works are dedicated to understanding and detecting hateful memes.
The Hateful Meme Challenge~\cite{KFMGSFBLNZMVRLHCRASYSOPSP20} launched by Facebook encourages a series of multimodal detection frameworks~\cite{M20,SCG19,Z20} that identify hateful or offensive memes with both visual and linguistic modalities.

However, as the study~\cite{SAADMFHSNC22} shows, identity attack, as an important type of hate, is under-explored by previous work.
Zannettou et al.~\cite{ZFBB20} find that the Happy Merchant meme enjoys substantial popularity on 4chan and Gab.
Some work~\cite{GZ22} studies hate speech and hateful imagery separately with image-text contrastive pre-trained models.
Targeting antisemitism and islamophobia, they first detect hateful textual phrases and then use the pre-trained CLIP to retrieve memes that are highly similar to hateful phrases.

\mypara{Meme evolution}
Research works on meme evolution~\cite{B11,ZCBCSSS18,DMCR18} have a preference for finding out how memes evolved and mutate variations over a period of time.
Bauckhage et al.~\cite{B11} study the temporal dynamics of 150 memes collected from Google Insights and three social bookmarking services, showing that user communities reflect different interests/behaviors of different memes.
Zannettou et al.~\cite{ZCBCSSS18} provide a large-scale assessment of meme popularity.
With the help of perceptual hashing (phash) and clustering techniques, they detect groups of memes and trace meme variations.
Dubey et al.~\cite{DMCR18} adopt a different solution to understand meme evolution and propagation.
They extract both visual and textual features from the same meme image and concatenate them into a new feature, which serves as the meme representation.
Leveraging a set of pre-selected template memes, they perform clustering (KNN) on the meme representations and retrieve the new variations.

% ----------------------------------------------------
\section{Discussion \& Conclusion}
\label{section:conclusion}
% ----------------------------------------------------

This paper presented a framework for understanding and analyzing hateful memes, with a particular focus on identifying variants of hateful memes and the images that are influencing the creation of these memes.
In particular, using a dataset obtained from 4chan's /pol/ and OpenAI's CLIP model that encapsulates semantic regularities in its generated embeddings, we identify the contents of hateful targets and perform a systematic analysis on the evolution of hateful memes, with a focus on antisemitic memes (i.e., the Happy Merchant meme).
Our analysis shows the multi-faceted aspect of the generation and evolution of hateful memes through the lens of the Happy Merchant meme.
In particular, using our framework, we identified 3.3K Happy Merchant variants shared on 4chan's /pol/.
At the same time, our findings show that 4chan users tend to create a large number of antisemitic Happy Merchant variants, as we find 80.0\% Happy Merchant variants for nationalities, religious, or political entities, 76.7\% variants for countries, 44.4\% for organizations, and 48.3\% for people.
We contribute toward this goal by proposing our framework that uses large-scale AI models that leverage the multimodal contrastive learning paradigm (such as OpenAI CLIP) to extract insights into the ecosystem of the generation and evolution of hateful memes.
We discuss the implications of our work by considering how our framework can be used for content moderation and tackling coordinated hate campaigns on the Web.

\mypara{Content moderation by combining AI and human moderators}
Online social media platforms such as Facebook and Twitter moderate content using automated tools (e.g., AI models) and human moderators that manually review content~\cite{fbmod}.
In the cases of harmful content (e.g., hateful symbols, child pornography, etc.), platforms like Facebook rely on a database of images/videos that share harmful content and hashing techniques to detect instances of harmful content in the wild~\cite{fb_hashing_moderation}.
This approach is not ideal for tackling the problem of harmful content on social media platforms as it does not generalize beyond the instances included in the existing database, and the generated hashes do not encapsulate the semantics of the images.
Due to these reasons, many instances of harmful content remain undetected on social media platforms or are only detected after many users reported the content.

Here, we propose using our framework for detecting variants of hateful content automatically, on a large scale, using the CLIP model and using content moderators to review the flagged content so that we minimize false positives generated by our framework.
For instance, given that social media platforms like Facebook already have a database of hateful symbols, they can leverage this ground truth and our framework to expand their database by identifying hateful variants.
When a new image is posted on the platform, our framework can assess whether the image is a hateful variant of any image that already exists in the database.
Then, the images will be presented to a human moderator that will determine if the image is indeed hateful.
Finally, the platform can take an automatic moderation action based on their existing hashing techniques for all confirmed hateful variants (identified by our framework and manually assessed by the moderators).
By employing our framework, we argue that social media platforms can improve their moderation workflow in a way that will have increased coverage in detecting and moderating emerging hateful variants.

\mypara{Coordinated hate campaigns} 
Our framework can play a significant role in identifying and mitigating coordinated hate campaigns~\cite{MSBCKLSS19} that unfold on the Web.
Coordinated hate campaigns involve the generation of new variants of hateful memes that start spreading on the Web to disseminate hateful ideologies targeting a specific individual/community.
Under such scenarios, our framework can be leveraged to identify new targets of hateful content using the clustering and hate assessment pipeline presented in \autoref{section: clustering}.
In this way, content moderators can quickly identify individuals or communities that might be the targets of orchestrated hate campaigns and take moderation interventions to mitigate the problem (e.g., post deletions or user bans/blocks~\cite{JBYB21,JGBG18} or soft moderation interventions~\cite{Z21,CJBG22}).
Additionally, social media platforms can use our framework to identify new variants of hateful content and undertake a temporal analysis (see \autoref{section: evolution}) to automatically identify spikes in the appearances of emerging hateful variants (i.e., a hateful variant appearing many times over a short time period).
By combining our target identification and hateful variant identification framework, we argue that social media platforms can promptly limit the effects of orchestrated hate campaigns.

\mypara{Limitations}
Our work has some limitations.
First, we demonstrate the application of our framework primarily using the Happy Merchant meme as a case study and focus on a single fringe social media platform (i.e., 4chan's /pol/).
Despite this limitation, we anticipate using our framework to generalize to new datasets from other social media platforms due to the great generalizability of large-scale AI models like OpenAI's CLIP model~\cite{RKHRGASAMCKS21}.
Indeed, by running our framework on other memes (e.g., Pepe the Frog, see \autoref{appendix: case_study_pepe} in the Appendix), we show that our framework generalizes beyond the Happy Merchant case study.
Second, our framework for identifying variants and influencers of hateful memes generates false positives that need to be considered carefully (see \autoref{section: evolution}), highlighting the need to keep humans in the loop when moderating content.
Despite this limitation, we argue that our framework can assist in understanding the evolution of hateful memes and help in moderating them.

\mypara{Future work} As part of our future work, we aim to apply our framework to understanding the evolution of other hateful memes on different platforms (e.g., misogynistic memes on Reddit).
Also, we aim to design and implement a dashboard that visualizes variants of hateful memes and their evolution over time, hence assisting moderators and journalists in understanding and mitigating hateful phenomena.

% ----------------------------------------------------
\section*{Acknowledgment}
% ----------------------------------------------------

We thank the Rewire team for providing us access to their API.
We also thank the anonymous reviewers and our shepherd for their invaluable comments and feedback that helped us improve our manuscript.
This work is partially funded by the Helmholtz Association within the project ``Trustworthy Federated Data Analytics'' (TFDA) (funding number ZT-I-OO1 4).

% ----------------------------------------------------
\begin{small}
\bibliographystyle{plain}
\bibliography{refs,normal_generated_py3}    
\end{small}
% ----------------------------------------------------

% ----------------------------------------------------
%\newpage
\appendix
\section{Appendix}
\label{appendix:appendix}
% ----------------------------------------------------

% ----------------------------------------------------
\subsection{Fine-tuning CLIP}
\label{appendix:finetuning}
% ----------------------------------------------------

We follow the standard training setup used by Radford et al.~\cite{RKHRGASAMCKS21} except for the learning rate and training epochs.
We decrease the original learning rate (5e-4) to 1e-6, where we observe a steady decrease in the loss value.
The original learning rate causes the model to be susceptible to gradients explosion when fine-tuned on our 4chan dataset.
Considering the training efficiency on a large amount of data, instead of setting a specific number of epochs, we monitor the loss changes over batch iterations.
We terminate fine-tuning at 60,000 iterations (approximately 3 epochs) after the loss value converges.

To evaluate the effectiveness of fine-tuning, we compare the performance of the pre-trained and fine-tuned CLIP models in a way similar to the evaluation of recommendation systems, as recommending is one of the most important applications of CLIP models and is critical for our following analysis.
Concretely, we randomly select 10,000 image-text pairs from both the training and the testing data, respectively, and consider them as two self-labeled datasets.
The performance difference between the seen and unseen data will indicate the generalizability of the CLIP model.
For every image, we calculate the similarities with all texts in the self-labeled dataset and select its top-$k$ results.

\begin{figure}
\centering
\includegraphics[width=\columnwidth]{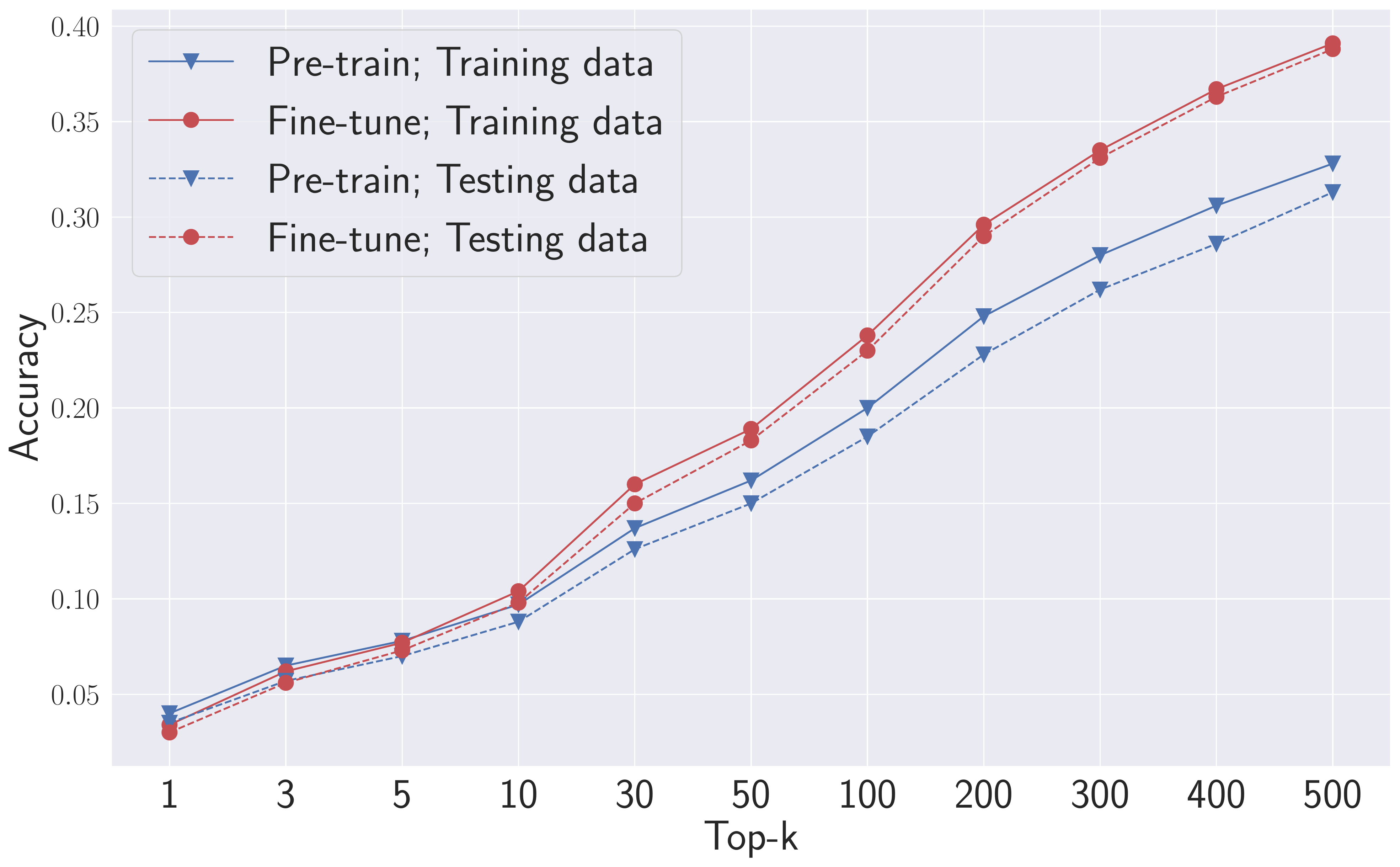}
\caption{Evaluation results of the pre-trained and the fine-tuned CLIP model on both the training and testing data.
Compared with the pre-trained model, the fine-tuned one presents a higher accuracy when it recommends a larger number of sentences for each image. 
The evaluation difference between the training and testing data is trivial.}
\label{figure:evaluation_CLIP}
\end{figure}

If the original text of a given image is contained in top-$k$ recommended results, we then consider the recommendation successful.
\autoref{figure:evaluation_CLIP} displays the recommendation accuracy between the pre-trained and fine-tuned CLIP models as top-$k$ increases.
For both the training and the testing data, the fine-tuned CLIP shows a 0.03-0.07 improvement in accuracy compared to the pre-trained CLIP when increasing top-$k$ from 50 to 500.
Furthermore, the minor evaluation difference between the training and testing data (i.e., the gap between red and blue lines) demonstrates the strong generalization ability of CLIP models.

% ----------------------------------------------------
\subsection{Probing CLIP's Semantic Regularities}
\label{appendix:probing}
% ----------------------------------------------------

Here we probe the intrinsic reason why semantic regularities generally exist in CLIP embeddings.
As linguistic semantic regularities on word embeddings have been studied extensively~\cite{CCP20}, we focus on visual semantic regularities.
Recall the image backbone in our fine-tuned CLIP adopts the Vision Transformer (ViT) architecture~\cite{DBKWZUDMHGUH21}, in which we later found that visual semantic regularities exist in image embeddings.  
We presume two possible reasons that contribute to the observed property: the model's image encoder architecture and the training approach (contrastive with image-text pairs vs. contrastive with image-image pairs).

We conduct a controlled experiment to find out which factor contributes the most.
Regarding the image encoder architecture, besides the CLIP-ViT we used before, we also fine-tune a CLIP-ResNet, a ResNet, and a ViT on the same 4chan dataset.
The model architecture of Vision Transformer (ViT) is substantially different from the Residual Neural Network (ResNet) where the former leverages attention blocks while the latter uses CNN blocks.
Note that the CLIP-ResNet is a variant model architecture pre-trained by CLIP and we fine-tune it in the same image-text contrastive way as CLIP.
The ResNet and ViT are trained from scratch in the image-image contrastive way (SimCLR~\cite{CKNH20}) due to the absence of supervised labels.

We then construct 10 influencer-variant image pairs of Happy Merchant, where the variants are the retrieved images from GPE entity category in \autoref{figure:evolution_image_text}.
Instead of using textual influencers, we search for the image influencers as introduced in \autoref{subsection: visual_semantic_regularities}.
Given the Happy Merchant image and the influencer image, we ask each model to return the top-3 images that are the closest to the summation embedding.
Comparing the returned images with the ground-truth variant, we consider the visual semantic relation is successfully captured if one of the returned images is semantically the same as the ground-truth variants.
We evaluate the accuracy of visual semantic capturing in each model.
As displayed in \autoref{table:eval_semantic}, CLIP-based models have a large percentage of variants that have the observed semantic regularities.
Thus, we believe the image-text contrastive training approach contributes more to the visual semantic regularities than the model architecture design.
We further conjecture the reason for CLIP's visual semantic regularities is that the connection built between images and texts drives the image encoder to learn the inherent semantic topology of texts.

\begin{table}[!t]
\centering
\caption{Evaluation of visual semantic capturing in different models. 
We construct 10 image pairs as ground truth and manually annotate the accuracy of semantic capturing.}
\label{table:eval_semantic}
\setlength{\tabcolsep}{0.25em}
\scalebox{0.9}{
\begin{tabular}{lccc}
\toprule
Model& Architecture & Training Type & Accuracy \\
\midrule
CLIP-ViT    & Transformer & image-text contrastive  & 10/10 \\
CLIP-ResNet & CNN         & image-text contrastive  & 7/10 \\
ViT         & Transformer & image-image contrastive &  1/10 \\
ResNet      & CNN         & image-image contrastive &  0/10 \\
\bottomrule
\end{tabular}
}
\end{table}

\begin{figure*}[!t]
\centering
\includegraphics[width=\textwidth]{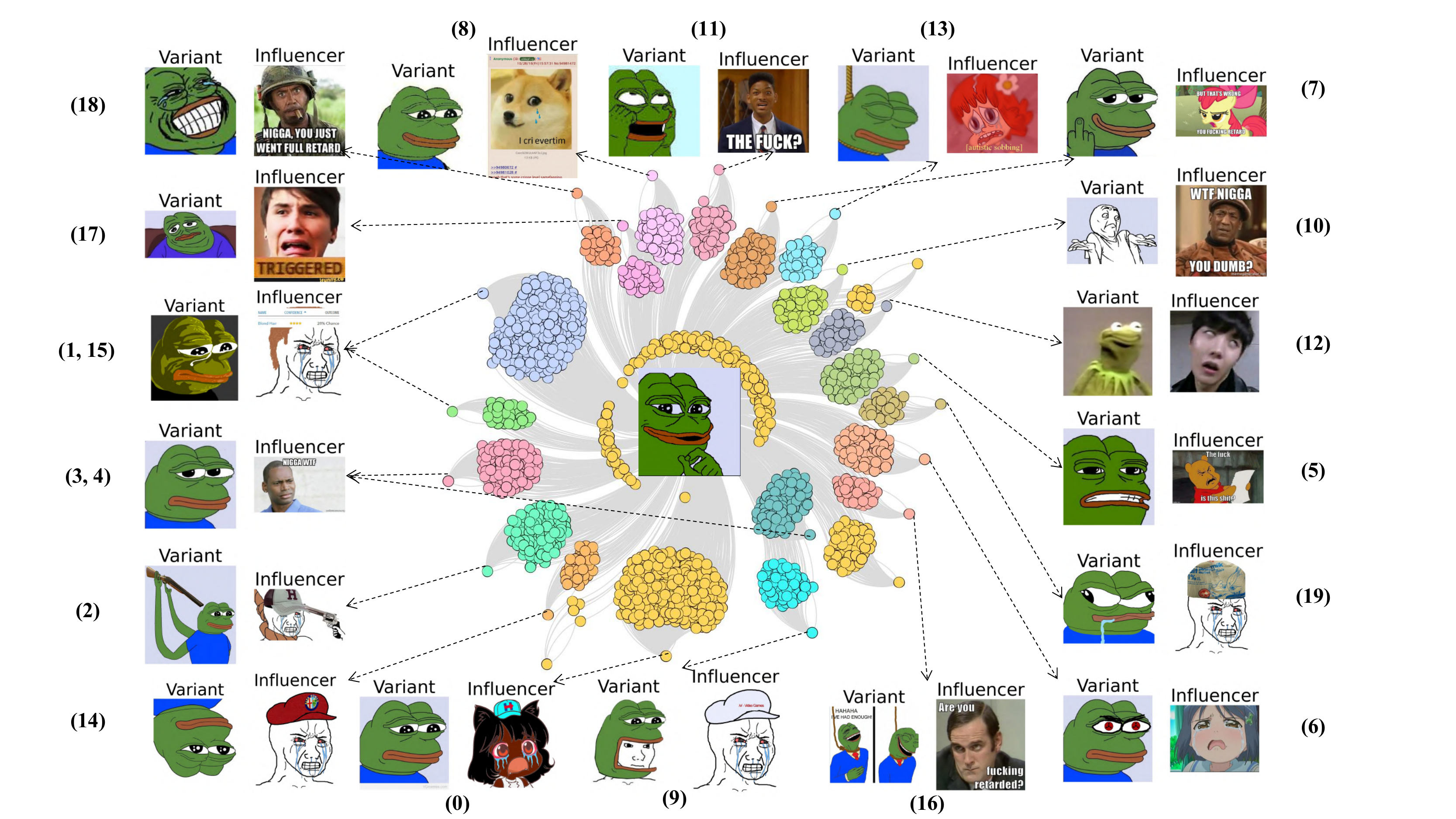}
\caption{The top-20 communities in the ecosystem of Pepe the Frog. 
Colors differentiate the communities. 
We annotate each community with two images: a variant on the left and its potential influencer on the right.}
\label{figure:pepe_evolution_image_image}
\end{figure*}

% ----------------------------------------------------
\subsection{More Case Study Results}
\label{appendix: case_study_pepe}
% ----------------------------------------------------

Here, we present our evolution analysis on a different meme to demonstrate our framework's generalizability in studying memes' evolution (beyond the Happy Merchant meme case study presented in the main text).
To do this, we focus on the Pepe the Frog meme, which is one of the most popular memes in 4chan~\cite{ZCBCSSS18} and is included in hate symbols by the Anti-Defamation League~\cite{ADL}.
We apply our two pipelines, namely, extracting visual semantic regularities and visual-linguistic semantic regularities, to identify Pepe the Frog variants.

\mypara{Visual semantic regularities}
Following the same threshold selection principle, we set [0.93, 0.95] as the similarity range to identify the Pepe the Frog variants and [0.89 0.96] as the range to recognize the image influencers.
We detected 6,357 pairs of variants and top-1 influencers, from which we randomly annotated 100 pairs and observed 94\% of variants and 37\% of influencers have been correctly identified.
\autoref{figure:pepe_evolution_image_image} displays the top-20 communities of Pepe the Frog variants.
Compared to the Happy Merchant case, the performance in identifying variants is better (an increase of 16\%), while it is weaker in recognizing influencers (a decrease of 16\%).
One of the potential reasons is that many Pepe the Frog variants in the top-20 communities do not have apparent external elements that can be identified.
On the contrary, the Happy Merchant variants are often fused with people, Nazi symbols, and other memes that are easier to recognize.
Finally, we look at whether we can increase the performance of our Influencer identification procedure by considering the top-10 similar images instead of top-1.
Here, we manually check all top-10 similar images to identify if any of the ten images can be classified as an influencer.
We find that by looking into the top-10 images, we can increase the performance by 10\% (overall influencer identification of 47\%).

\begin{figure*}[!t]
\centering
\includegraphics[width=\textwidth]{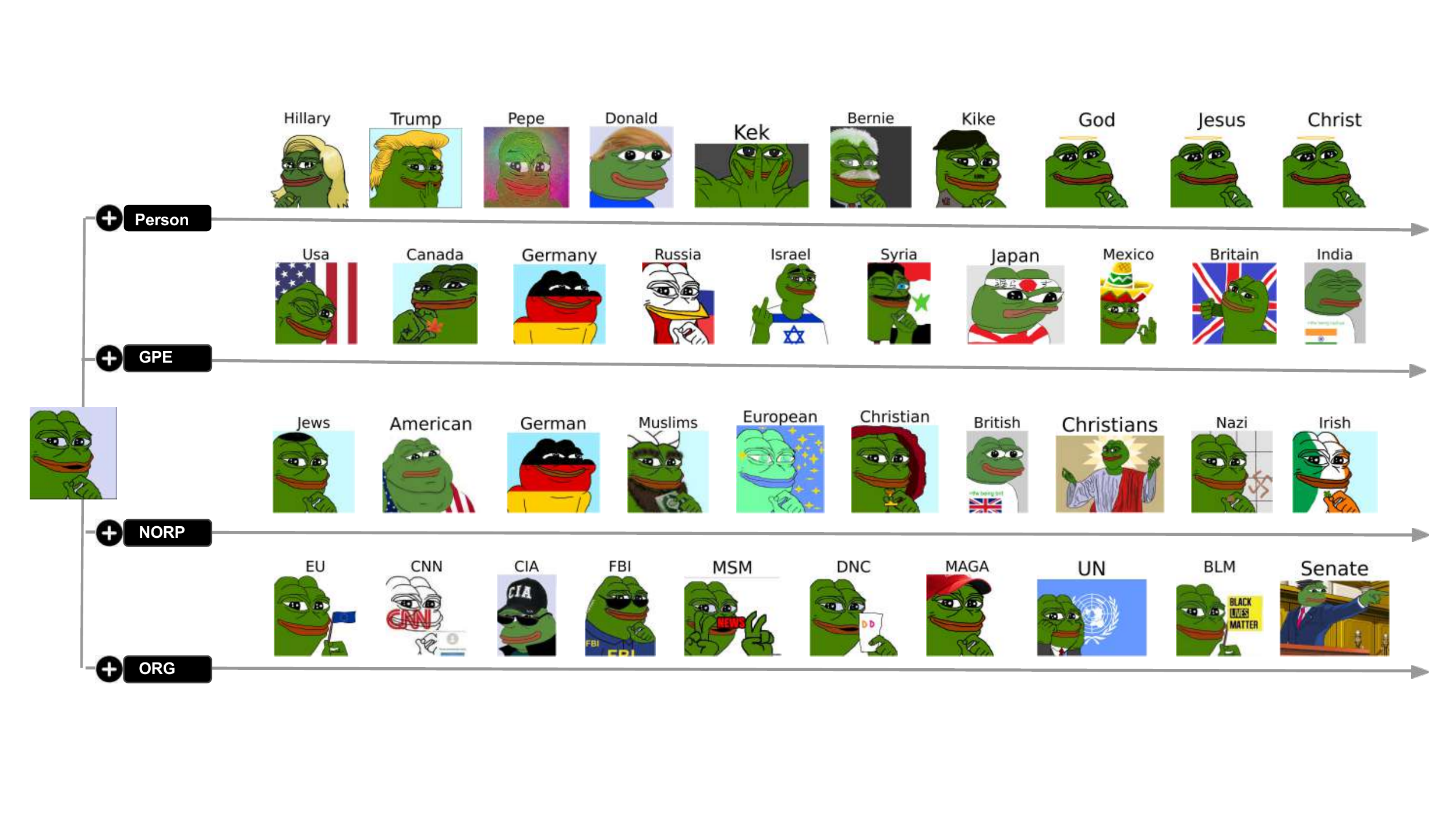}
\caption{Pepe the Frog variants influenced by the four types of entities. For each type of entity, we visualize 10 examples for each type.}
\label{figure:pepe_image_text}
\end{figure*}

\mypara{Visual-linguistic semantic regularities}
Here, we use the same four categories (People, GPE, NORP, and ORG) of named entities used in the Happy Merchant case study to identify variants of the Pepe the Frog meme using visual-linguistic semantic regularities.
The annotation result shows 37.9\% of entities in People, 83.3\% in GPE, 80.0\% in NORP, and 63.0\% in ORG have merged variants.
\autoref{figure:pepe_image_text} displays ten variant examples for each entity category.
There are several insights when comparing the case study results of Pepe the Frog and Happy Merchant.
They are both prone to meddle with political entities, e.g., for GPE and NORP entities, we have the greatest possibilities to find merged variants.
Also, the variants of Pepe the Frog influenced by the same entity can sometimes be more diverse than the Happy Merchant case, e.g., the variants in \autoref{figure:pepe_image_text} influenced by Trump and Donald.

\begin{figure*}[ht]
\centering
\begin{subfigure}{0.49\textwidth}
\includegraphics[width=1\linewidth]{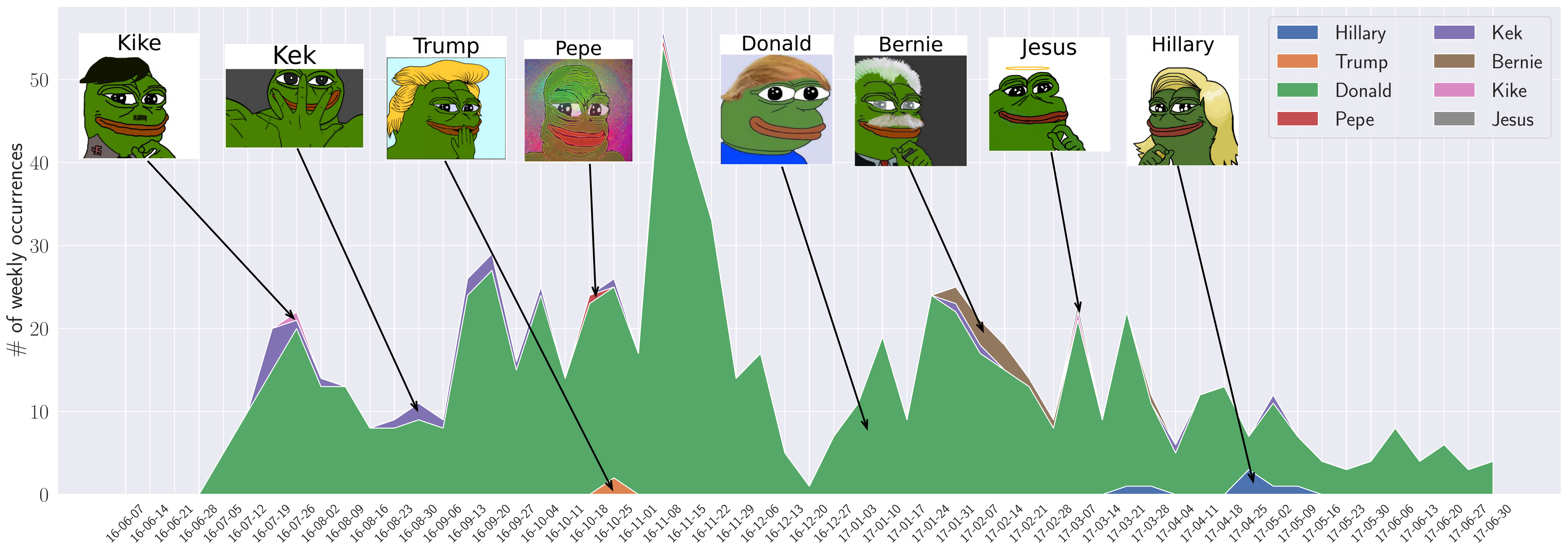}
\caption{People}
\label{figure:pepe_people} 
\end{subfigure}
\begin{subfigure}{0.49\textwidth}
\includegraphics[width=1\linewidth]{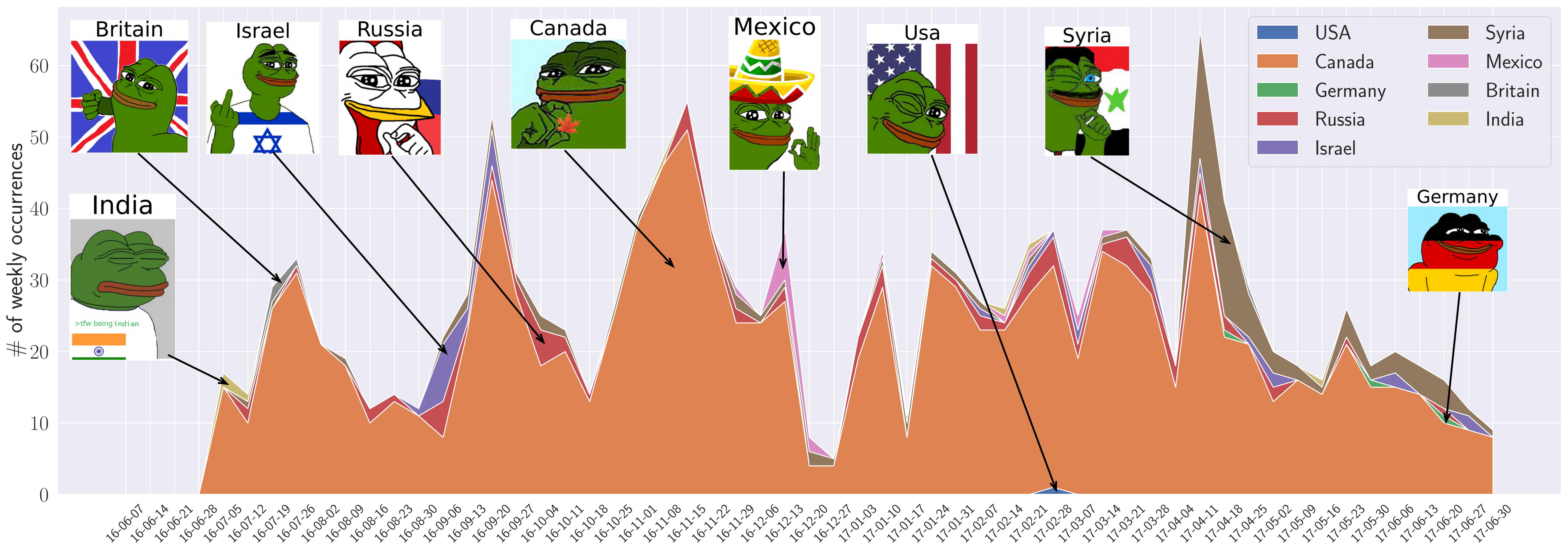}
\caption{GPE}
\label{figure:pepe_gpe}
\end{subfigure}
\medskip
\begin{subfigure}{0.49\textwidth}
\includegraphics[width=1\linewidth]{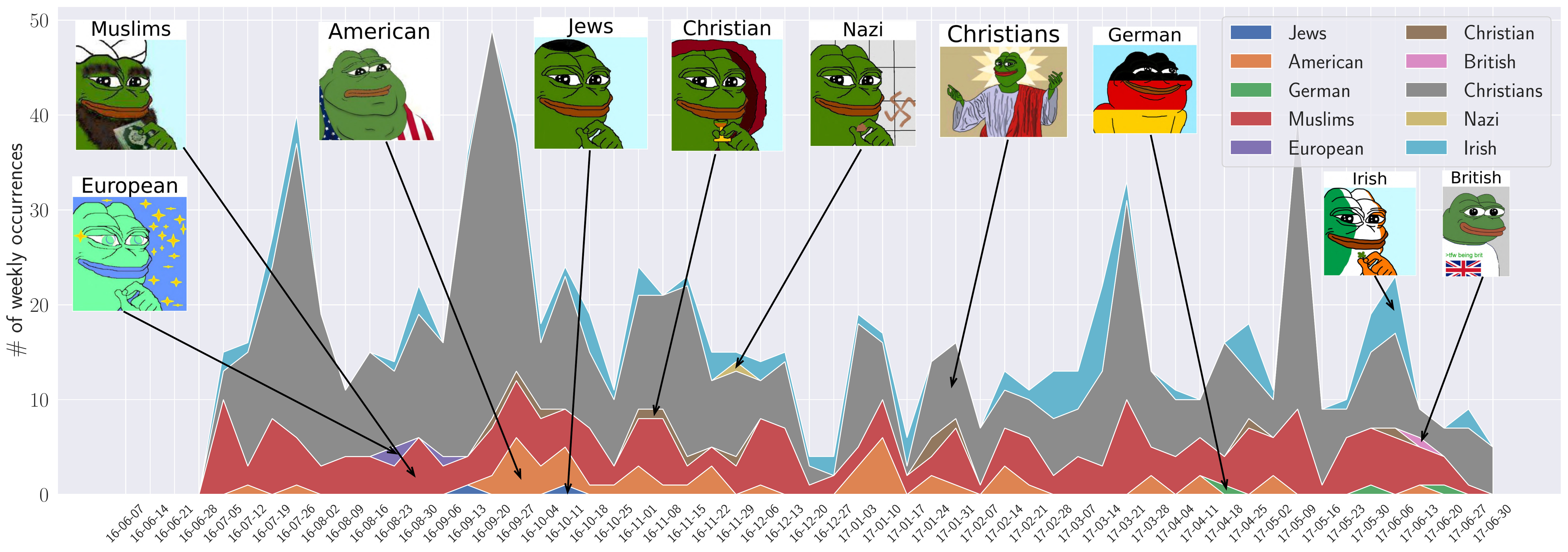}
\caption{NORP}
\label{figure:pepe_norp} 
\end{subfigure}
\begin{subfigure}{0.49\textwidth}
\includegraphics[width=1\linewidth]{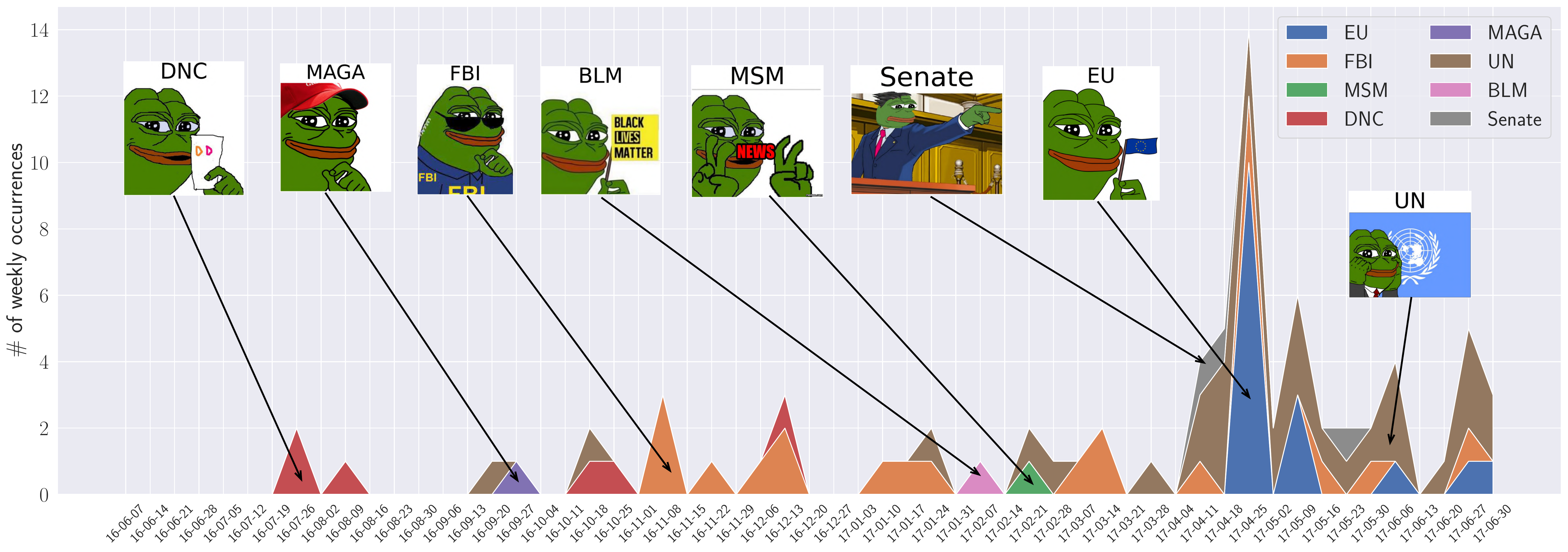}
\caption{ORG}
\label{figure:pepe_org}
\end{subfigure}
\caption{Number of posts including Pepe the Frog variants per week.}
\label{figure:temporal_pepe}
\end{figure*}

\mypara{Temporal analysis}
We conduct a temporal analysis on the Pepe the Frog variants identified above following the same procedure as the Happy Merchant case.
\autoref{figure:temporal_pepe} shows the temporal change in the number of posts that include Pepe the Frog variants.
Compared to the temporal change in the Happy Merchant case, we observe some commonalities with the main spreading period of different hateful variants.
For example, both hateful memes merged with the entity MAGA are shared in September 2016 (\autoref{figure:org}, \autoref{figure:pepe_org}); and variants of Mexican (or Mexico) both appear in November 2016 (\autoref{figure:norp}, \autoref{figure:pepe_gpe}).
We also observe some differences, e.g., the ``Pepe-Donald'' variant is more popular than the ``Pepe-Hillary'' variant, in contrast to the Happy Merchant case (\autoref{figure:people}, \autoref{figure:pepe_people}).
This may indicate that 4chan users have different inclinations when fusing hateful memes with certain politicians.
Also, when considering the GPE (see \autoref{figure:pepe_gpe}), we find that 4chan users consistently share the Canada variants of Pepe the Frog, which likely indicates that 4chan users share meme variants to disseminate anti-Canadian sentiments.

% ----------------------------------------------------
\end{document}